\documentclass[sigconf]{acmart}

\AtBeginDocument{%
  \providecommand\BibTeX{{%
    \normalfont B\kern-0.5em{\scshape i\kern-0.25em b}\kern-0.8em\TeX}}}

\copyrightyear{2026}
\acmYear{2026}
\setcopyright{cc}
\setcctype{by}
\acmConference[ASSETS '26]{The 28th International ACM SIGACCESS Conference on Computers and Accessibility}{October 25--28, 2026}{Vila Nova de Gaia, Portugal}
\acmBooktitle{The 28th International ACM SIGACCESS Conference on Computers and Accessibility (ASSETS '26), October 25--28, 2026, Vila Nova de Gaia, Portugal}
\acmDOI{10.1145/3797867.3829019}
\acmISBN{979-8-4007-2521-0/2026/10}

\usepackage[commandnameprefix=ifneeded]{changes} 
\usepackage{todonotes}
\usepackage[skins]{tcolorbox}

\presetkeys{todonotes}{size=\small}{}

\newlength{\CommentTargetMarginWidth}
\newif\ifCommentAutoAdjust
\CommentAutoAdjusttrue
\newif\ifCommentMarginAdjusted
\CommentMarginAdjustedfalse
\newcommand{\EnsureCommentMargin}{%
	\ifCommentAutoAdjust
		\ifCommentMarginAdjusted
		\else
			\ifdim\marginparwidth<\CommentTargetMarginWidth
				\setlength{\marginparwidth}{\CommentTargetMarginWidth}%
			\fi
			\global\CommentMarginAdjustedtrue
		\fi
	\fi
}

\colorlet{BaseHighlightColor}{yellow!22}
\colorlet{JYHighlightColor}{blue!14}
\colorlet{SYHighlightColor}{red!16}
\colorlet{SCTHighlightColor}{green!18}
\colorlet{AZHighlightColor}{orange!20}
\colorlet{OMHighlightColor}{teal!18}
\colorlet{WJHighlightColor}{violet!18}

\colorlet{JYTodoColor}{blue!40}
\colorlet{SYTodoColor}{red!40}
\colorlet{SCTTodoColor}{green!40}
\colorlet{AZTodoColor}{orange!40}
\colorlet{OMTodoColor}{teal!35}
\colorlet{WJTodoColor}{violet!35}

\newcommand{\CommentBlockHighlight}[2]{%
  \noindent\begin{tcolorbox}[
    colback=#1,
    colframe=#1!55!black,
    boxrule=0pt,
    left=4pt,
    right=4pt,
    top=3pt,
    bottom=3pt,
    sharp corners,
    before skip=6pt,
    after skip=6pt
  ]
    #2
  \end{tcolorbox}%
}

\newcommand{\CommentInlineHighlight}[2]{%
  \csname tcbox\endcsname[colback=#1,
         colframe=#1!55!black,
         boxrule=0pt,
         left=2pt,
         right=2pt,
         top=1pt,
         bottom=1pt,
         sharp corners,
         on line]
        {#2}%
}

\newcommand{\AuthorAnnotation}[5]{%
  \EnsureCommentMargin
  \ifhmode
    \CommentInlineHighlight{#1}{#2}%
  \else
    \CommentBlockHighlight{#1}{#2}%
  \fi
  \csname todo\endcsname[color=#5]{#3: #4}%
  \ignorespaces
}

\usepackage{pifont}
\usepackage{graphicx}
\usepackage{natbib}
\usepackage{enumitem}
\usepackage{tabularx} \usepackage{caption}
\usepackage{subcaption}
\usepackage{ragged2e} 
\usepackage{placeins} 
\usepackage{tikz}
\newcolumntype{Y}{>{\RaggedRight\arraybackslash}X}

\usepackage{colortbl}   
\definecolor{okSuccess}{HTML}{009E73}  
\definecolor{okPartial}{HTML}{E69F00}  
\definecolor{okFailure}{HTML}{D55E00}  
\definecolor{okDivider}{HTML}{444444}
\newcommand{\cS}{\cellcolor{okSuccess}\textcolor{white}{\ding{51}}}   
\newcommand{\cP}{\cellcolor{okPartial}\textcolor{black}{$\boldsymbol{\sim}$}} 
\newcommand{\cF}{\cellcolor{okFailure}\textcolor{white}{\ding{53}}}

\begin{document}

\title{QUARTZ: Qualitative Understanding via Accessible Representation and Visualization}

\author{Omar Khan}
\email{mkhan259@illinois.edu}
\orcid{0009-0005-3209-3525}
\affiliation{%
    \department{Siebel School of Computing and Data Science}
    \institution{University of Illinois Urbana-Champaign}
    \city{Urbana}
    \state{Illinois}
    \country{USA}
}

\author{JooYoung Seo}
\email{jseo1005@illinois.edu}
\orcid{0000-0002-4064-6012}
\affiliation{%
    \department{School of Information Sciences}
    \institution{University of Illinois Urbana-Champaign}
    \city{Champaign}
    \state{Illinois}
    \country{USA}
}

\renewcommand{\shortauthors}{Khan and Seo}

\begin{abstract}

Qualitative data visualizations -- concept maps, network graphs, Sankey diagrams, and coding stripes -- are integral to research practice, yet remain entirely inaccessible to blind and low-vision (BLV) researchers. 
While visualization has seen advanced multimodal solutions for quantitative charts, qualitative visualizations, and their non-linear, semantically rich structures have received no attention. We present QUARTZ, a web-based system that provides screen-reader-accessible, multimodal representations of qualitative data visualizations. 
Using the Rapid Iterative Testing and Evaluation (RITE) method, we conducted a formative user study with 8 BLV participants who completed 12 tasks across four visualization types. 
Our findings expose accessibility barriers unique to qualitative visualizations, non-linear navigation breakdowns and semantic comprehension gaps absent from quantitative chart research, and document how iterative co-design with BLV users progressively addressed them across sessions.
We contribute empirical evidence and design guidelines for an underexplored visualization domain, advancing the infrastructure BLV researchers need to participate independently in qualitative inquiry.
\end{abstract}




\maketitle

\section{Introduction}
\label{sec:introduction}

Concept maps, network graphs, Sankey diagrams, and coding stripes are not decorative outputs of qualitative research -- they are analytic instruments through which researchers identify patterns, develop theory, and communicate findings~\cite{milesQualitativeDataAnalysis2014, hendersonVisualizingQualitativeData2013}. Qualitative data analysis has grown in its scientific leverage in recent decades~\cite{thelwallResearchQualitativeData2021}, and major QDA software packages treat visualization as integral to the workflow, embedding visual outputs at multiple stages of the research process. Yet for blind and low-vision (BLV) researchers, every one of these visualizations is inaccessible.

The consequences are concrete. QDA tools rely on color-coded themes, drag-and-drop arrangement, and hover-based inspection, affordances that are fundamentally incompatible with screen reader interaction~\cite{milesQualitativeDataAnalysis2014, aishwaryaPerformingQualitativeData2022}. In a mixed-methods study with 57 survey respondents and 15 interviewees, Khan et al. found that data analysis and visualization represent the most challenging research stage for BLV researchers, and nearly one-fifth of BLV researchers reported being unable to independently evaluate visual outputs, instead delegating these tasks to sighted colleagues~\cite{khan_i_2026}. This is a pragmatic workaround that creates delays, limits professional development, and undermines epistemic autonomy~\cite{khan_i_2026}.

The accessible visualization community has responded to analogous problems in quantitative charting. Systems such as MAIDR~\cite{seoMAIDRMakingStatistical2024, seoMAIDRMeetsAI2024}, Umwelt~\cite{zongUmweltAccessibleStructured2024}, VoxLens~\cite{sharifShouldSayDisabled2022}, Chart reader~\cite{thompsonChartReaderAccessible2023}, and Olli~\cite{blancoOlliExtensibleVisualization2022} provide multimodal access to Cartesian-based plots, such as bar plots, scatter plots, heat maps, box plots, and line charts through combinations of sonification, textual description, Braille output, and keyboard navigation. These systems have demonstrated that BLV users can accurately interpret data visualizations when appropriate non-visual representations are provided.


However, qualitative data visualizations differ from their quantitative counterparts in fundamental ways: they encode semantic relationships rather than numerical magnitude, employ non-linear relational structures rather than ordered axes, and vary widely in form depending on the analytic tradition and research question. There is limited prior work that provides accessible representations of qualitative visualization types and empirical evidence on how BLV users navigate, interpret, or construct meaning from these representations~\cite{blancoOlliExtensibleVisualization2022}. 

This gap has direct consequences for epistemic equity in research. BLV researchers who conduct qualitative studies:  interviewing participants, transcribing data, and developing codebooks, are systematically excluded from the visual analytic tools that their non-BLV colleagues use to make sense of that same data. The result is a forced division of labor driven by tool inaccessibility, not by competence. Addressing this gap is necessary to enable the full and independent participation of BLV researchers in qualitative inquiry.

We present QUARTZ (Qualitative Understanding via Accessible Representation and Visualization), a web-based system that provides accessible, multimodal representations of qualitative data visualizations for BLV users. QUARTZ supports four visualization types: concept maps, network graphs, Sankey diagrams, and coding stripes, through screen-reader-accessible structured navigation, sonification, and textual summaries. We evaluate the system's exploration and comprehension capabilities through moderated, task-based usability testing sessions with 8 BLV participants who completed 12 tasks across the four visualization types using a Rapid Iterative Testing and Evaluation (RITE)~\cite{RapidIterativeTest2005} approach that allowed us to surface and resolve accessibility barriers between sessions rather than documenting them passively.


Our study is guided by four research questions:

\begin{enumerate}
    \item[\textbf{RQ1)}] How can a visualization exploration system guide BLV users toward qualitative visualization types that appropriately match their data characteristics while constraining complexity to support non-visual comprehension?
    \item[\textbf{RQ2)}] To what extent does QUARTZ enable BLV users to independently explore qualitative data visualizations compared to their current workflows involving general-purpose tools or sighted assistance?
    \item[\textbf{RQ3)}] How do BLV users experience the multi-modal feedback mechanisms in QUARTZ to support autonomy and control throughout the visualization exploration process?
    \item[\textbf{RQ4)}] How do BLV users assess visualization quality through non-visual evaluation mechanisms in QUARTZ?

\end{enumerate}


This paper makes the following contributions:

\begin{enumerate}
  \item \textbf{System:} We present QUARTZ, an accessible qualitative data visualization system, supporting four visualization types through multimodal interaction designed by, with, and for BLV users.
  \item \textbf{Empirical findings:} We provide empirical characterization of \textit{how} BLV users interact with qualitative data visualizations, identifying accessibility barriers unique to this visualization domain, including non-linear navigation challenges, semantic comprehension difficulties, and screen-reader-specific interaction conflicts that do not arise in quantitative chart accessibility.
  \item \textbf{Design knowledge:} We derive design guidelines for accessible qualitative visualization, grounded in barriers surfaced and resolved through iterative testing with BLV users. These guidelines address a gap in the accessible visualization literature, which has to date focused exclusively on quantitative chart types.
\end{enumerate}


The remainder of this paper is organized as follows. Section~\ref{sec:related_work} reviews related work on accessible data visualization, qualitative data analysis, and the accessibility barriers BLV researchers face. Section~\ref{sec:design_procedures_and_goals} describes our design procedures and the goals that emerged from our co-design process. Section~\ref{sec:system_implementation} presents the QUARTZ system architecture and user workflow. Section~\ref{sec:evaluation} details our RITE-based evaluation methodology. Section~\ref{sec:findings} reports our findings, organized by accessibility barriers, design iterations, user strategies, and perceived value. Sections~\ref{sec:discussion} and ~\ref{sec:conclusion} discuss implications for future work in accessible qualitative visualization.

\section{Related Work}
\label{sec:related_work}


\subsection{The Landscape of Accessible Data Visualization}
\label{subsec:rw_accessible_viz}

Accessible data visualization is not a single problem but a family of them, spanning disability communities with distinct perceptual and cognitive needs. ~\citet{marriott_inclusive_2021} issued a call to action for inclusive
visualization across disability groups, and recent work has begun answering
it beyond the blind and low-vision community: ~\citet{tran_discovering_2024} identified visualization designs that support people with ADHD, and ~\citet{choiVisualizingNonVisualEnabling2019} developed pipelines that transform existing charts for visually impaired readers. Within this broader landscape, the deepest infrastructure, including design frameworks, evaluation heuristics, and working systems, serves BLV users of quantitative statistical charts. ~\citet{kimAccessibleVisualizationDesign2021} mapped the design space across visual, auditory, and haptic modalities, and ~\citet{lundgardAccessibleVisualizationNatural2022} formalized what natural language descriptions
of charts should convey at four levels of semantic content.

These systems have collectively demonstrated several capabilities that BLV users need: multimodal exploration through synchronized Braille, text, sonification, and review channels~\cite{seoMAIDRMakingStatistical2024, seoMAIDRMeetsAI2024}; independent authoring of data representations without sighted assistance~\cite{zongUmweltAccessibleStructured2024}; conversational and voice-based querying of chart content~\cite{sharifVoxLensMakingOnline2022, seoMAIDRMeetsAI2024}; hierarchical and structured keyboard navigation~\cite{thompsonChartReaderAccessible2023, elavskyDataNavigatorAccessibilityCentered2023, srinivasanAzimuthDesigningAccessible2023}; touch- and speech-based exploration of node-link diagrams \cite{zhao_tada_2024} and
non-visual reasoning over linked data structures~\cite{wimer_nonvisual_2026}; tactile output for physical reading~\cite{chenTactileVegaLiteRapidly2025}; and natural-sound sonification as an alternative to synthetic tones~\cite{hoqueAccessibleDataRepresentation2023, zhaoDataSonificationUsers2008}. Additional work has addressed extensible screen reader support~\cite{blancoOlliExtensibleVisualization2022}, consistent cross-device multimodal interaction~\cite{srinivasanInChorusDesigningConsistent2020}, design dimensions for rich screen reader experiences~\cite{zongRichScreenReader2022}, alternative text effectiveness~\cite{jungCommunicatingVisualizationsVisuals2022}, and the broader call for inclusive visualization across disability types~\cite{marriott_inclusive_2021, choiVisualizingNonVisualEnabling2019}.

A consistent finding across this work is the heterogeneity of BLV users' assistive technology setups: different screen readers, operating systems, and navigation strategies produce different experiences with the same system~\cite{seoMAIDRMakingStatistical2024, fanAccessibilityDataVisualizations2023}. This reality demands iterative evaluation with real users, not reliance on heuristics alone.

The field has made substantial progress, but the majority of these capabilities have been demonstrated exclusively for \textit{quantitative} visualizations. These chart types share structural properties that make non-visual encoding tractable: they map numerical values to ordered positions along axes, represent data as discrete elements that can be sequentially enumerated, and follow standardized layout conventions. Qualitative data visualizations share none of these properties.

\subsection{Qualitative Data Visualization vs. Quantitative Data Visualization}
\label{subsec:rw_qual_viz}

\begin{figure*}[t]
    \centering
  
    \begin{subfigure}[b]{\columnwidth}
        \centering
        \includegraphics[width=\linewidth]{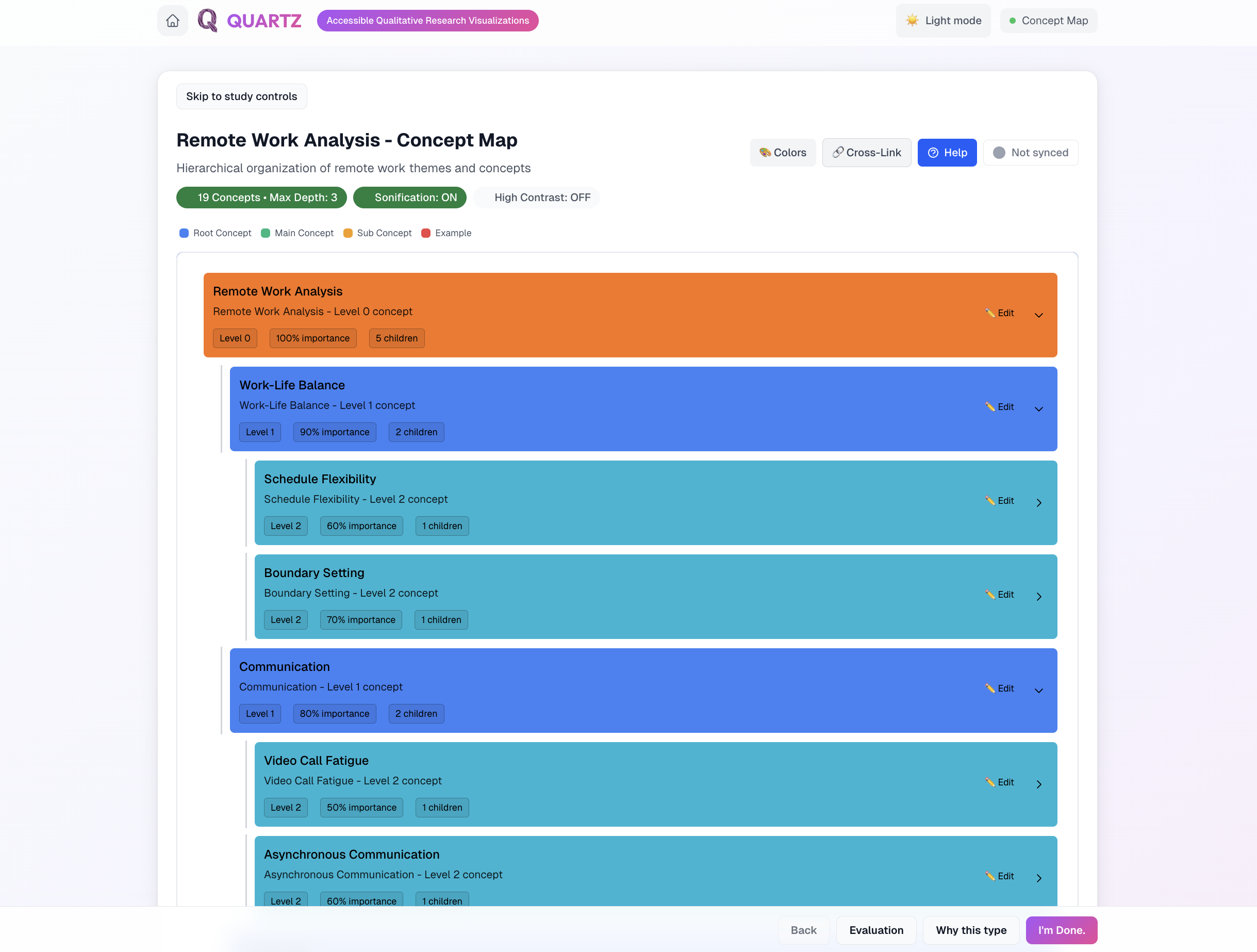}
        \caption{QUARTZ's concept map view.}
        \label{fig:sub1}
    \end{subfigure}
    \hfill
    \begin{subfigure}[b]{\columnwidth}
        \centering
        \includegraphics[width=\linewidth]{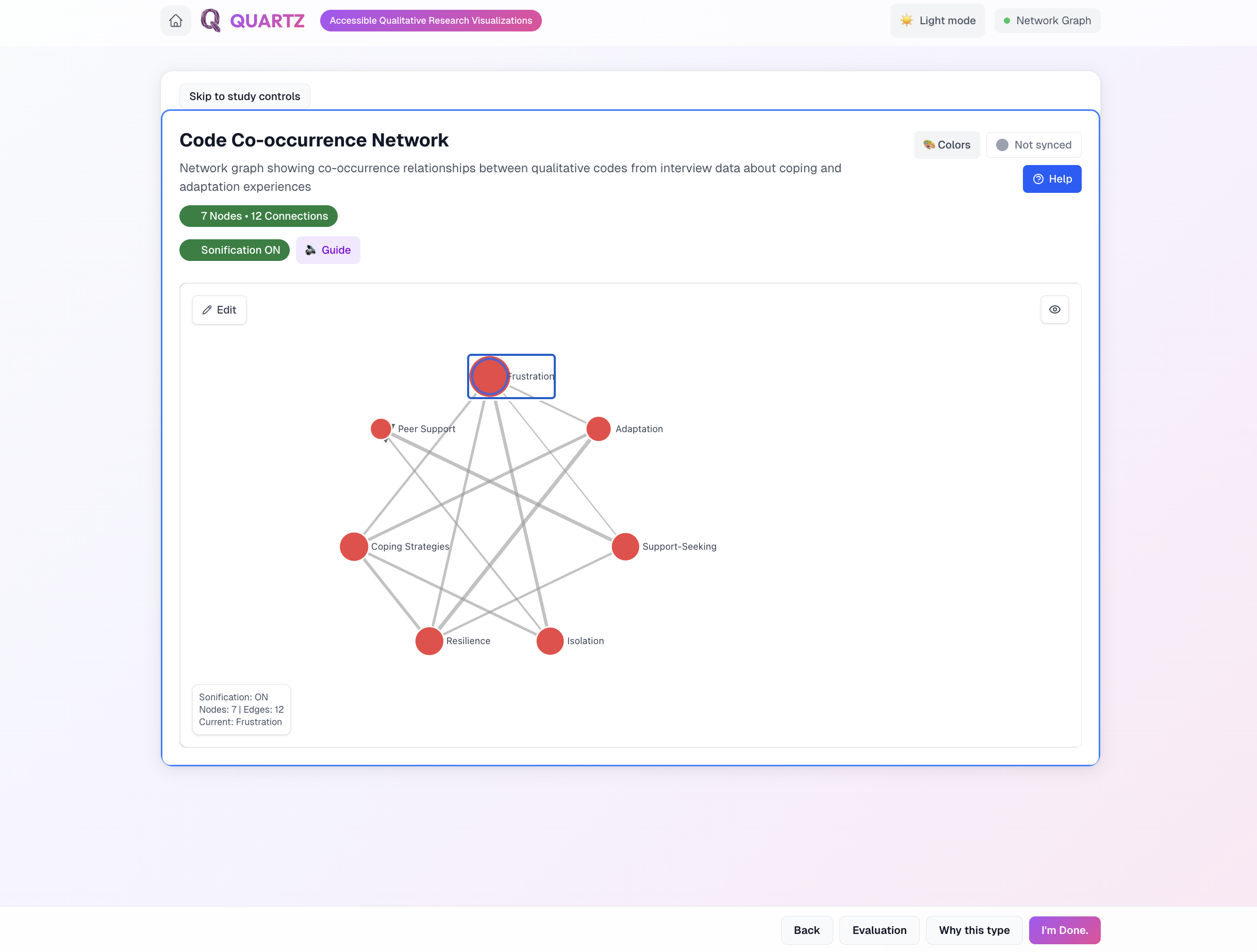}
        \caption{QUARTZ's network graph view.}
        \label{fig:sub2}
    \end{subfigure}

    \vspace{1em}

    \begin{subfigure}[b]{\columnwidth}
        \centering
        \includegraphics[width=\linewidth]{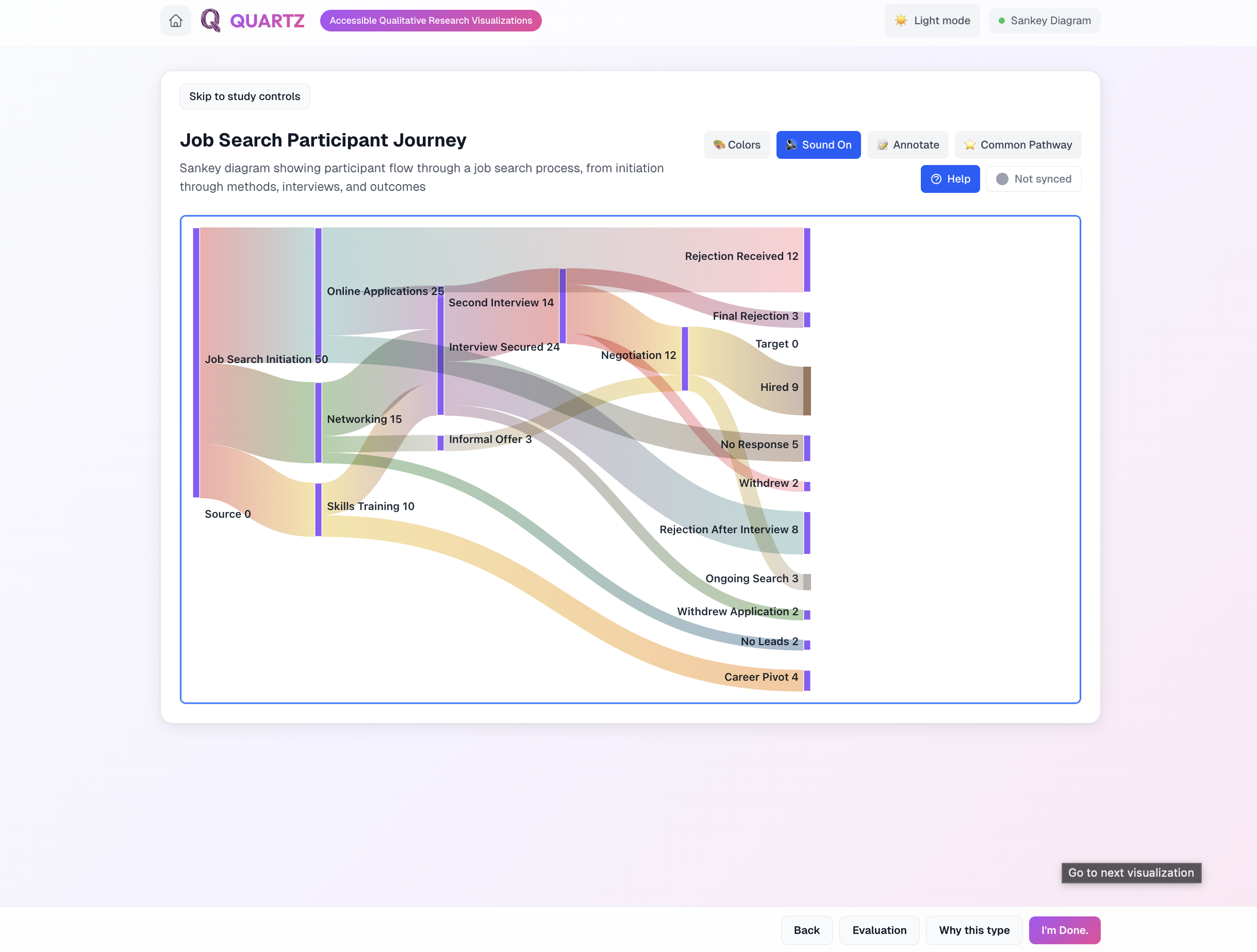}
        \caption{QUARTZ's Sankey diagram view.}
        \label{fig:sub3}
    \end{subfigure}
    \hfill
    \begin{subfigure}[b]{\columnwidth}
        \centering
        \includegraphics[width=\linewidth]{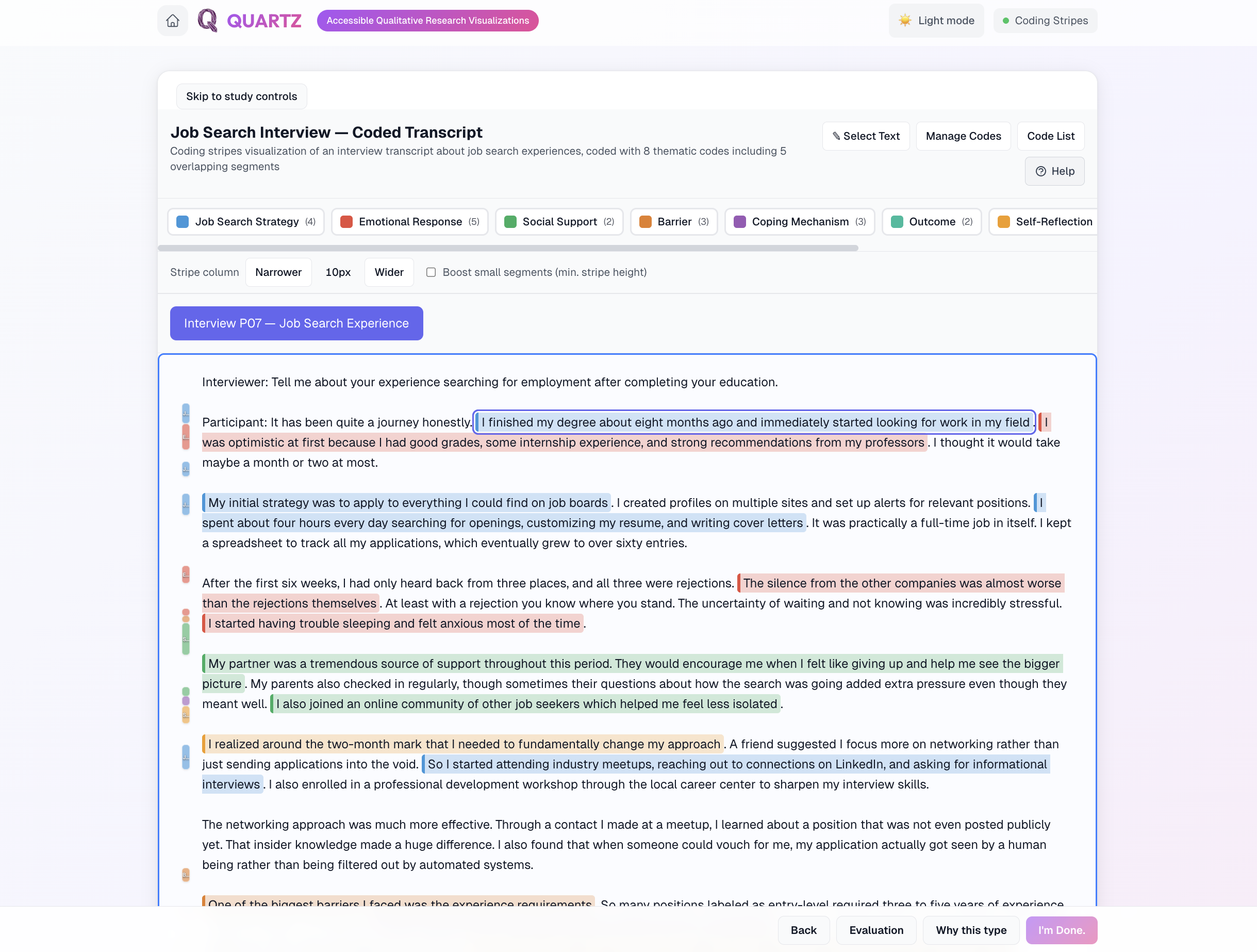}
        \caption{QUARTZ's coding stripes view.}
        \label{fig:sub4}
    \end{subfigure}

    \caption{QUARTZ's four visualization types.}
    \label{fig:four_images_main}
    \Description{Four screenshots of QUARTZ's visualization types arranged in a two-by-two grid. (a) Concept map showing hierarchical nodes with parent and child relationships. (b) Network graph showing nodes connected by weighted edges. (c) Sankey diagram showing flows between sequential stages. (d) Coding stripes showing text segments labeled with colored code stripes.}
\end{figure*}

Qualitative data visualization is not a variant of quantitative charting
with different labels. It constitutes a distinct visualization practice with
its own representational traditions~\cite{hendersonVisualizingQualitativeData2013, nelsonConversationTimeNew2022}, structural properties~\cite{pokornyNetworkAnalysisVisualization2018},
and methodological functions~\cite{decuypereVisualNetworkAnalysis2020, oanceaQualitativeNetworkAnalysis2017}.

Visualization serves as an analytic instrument in qualitative research, not merely a reporting tool. \citet{milesQualitativeDataAnalysis2014} describe qualitative data displays: matrices, networks, and flow charts, as tools for organizing and examining data that complement textual analysis. \citet{hendersonVisualizingQualitativeData2013} argue that qualitative visualization communicates findings that resist textual expression: temporal patterns, relational structures, the density and distribution of analytic categories. Several methodological traditions treat visualization as a method in its own right. \citet{decuypereVisualNetworkAnalysis2020} demonstrates visual network analysis as a qualitative method for investigating sociomaterial practices, using network diagrams as analytic objects rather than summary displays. \citet{pokornyNetworkAnalysisVisualization2018} provide a framework for representing thematic relationships as graph structures in psychology. \citet{oanceaQualitativeNetworkAnalysis2017} apply qualitative network analysis to trace the articulation of cultural value in research evaluation. 

New visualization forms continue to emerge: \citet{nelsonConversationTimeNew2022} adapted stream graphs for qualitative data, using Sankey-like temporal flows to track how conversational themes surface and recede. \citet{panagiotidouCommunicatingQualitativeUncertainty2022} investigated how qualitative uncertainty, a challenge with no quantitative parallel, can be communicated visually in the digital humanities.

The four visualization types QUARTZ supports, concept maps, network graphs, Sankey diagrams, and coding stripes (Figure~\ref{fig:four_images_main}), exemplify why qualitative visualizations demand different accessibility approaches. Two terminological clarifications: QUARTZ renders concept maps as a structured, list-like tree rather than the free-form node-link canvas the term sometimes connotes, a choice motivated by screen reader traversal order; and our coding stripes follow the QDA-software convention of colored code indicators rendered alongside coded text, as in NVivo, rather than instructional "coding strips" used in other literatures. Each type differs from quantitative charts in the following ways:

\begin{itemize}
  \item \textbf{Semantic encoding.} Nodes in a concept map represent thematic categories, not numerical values. Edges encode interpretive relationships ("is related to", "causes", "contradicts"), not mathematical functions. A node's meaning is inseparable from its qualitative content.
  \item \textbf{Non-linear topology.} Network graphs and concept maps are graph structures that may lack a natural reading order. A bar chart affords left-to-right axis traversal; a network graph \textit{may} afford hierarchical exploration, but this is not guaranteed. Graph readability depends on spatial layout~\cite{ghoniem_readability_2005}, which screen readers cannot convey.
  \item \textbf{Variable complexity.} A concept map may contain five nodes or fifty, with hierarchical nesting that varies by analytic depth. Unlike standardized chart types whose structural complexity is bounded by data dimensions, qualitative visualizations reflect the subjective scope of the researcher's coding scheme.
  \item \textbf{Interpretive subjectivity.} Two researchers analyzing the same data may produce structurally different concept maps, reflecting different analytic interpretations. Qualitative visualizations are products of judgment, not deterministic data transformations.
\end{itemize}

Sequential axis traversal, sonification of numerical magnitude, and data point enumeration -- the interaction paradigms that make quantitative charts accessible -- need to be uniquely considered for qualitative data. There is a need for increased design knowledge for such graphics, and limited insights currently exist.

\subsection{Barriers to Equitable Qualitative Research for BLV Researchers}
\label{subsec:rw_blv_researchers}

The accessibility barriers BLV researchers face in qualitative workflows have been increasingly documented, though have limited remedies. \citet{milesQualitativeDataAnalysis2014} identified that standard QDA tools, such as NVivo, ATLAS.ti, and MAXQDA, rely on visual affordances that are fundamentally incompatible with screen reader interaction: color-coded themes, drag-and-drop spatial arrangement, and hover-based inspection. BLV researchers develop workarounds, including delegating
visualization-dependent tasks to sighted colleagues, but these workarounds
erode autonomy rather than enabling it~\citep{khan_i_2026}. These barriers are
methodological as much as technical: Emara~\citep{emaraTalkingSameLanguage2025} showed that visual
impairment identity shapes rapport and recruitment in qualitative studies,
meaning accessibility conditions not only which tools BLV researchers can
use but how they conduct qualitative research at all. The tooling dimension,
however, remains the least addressed.

Khan and Seo exposed the scope of this exclusion through an explanatory sequential mixed-methods study with 57 BLV researchers (survey) and 15 BLV researchers (interviews), finding that data analysis and visualization emerged as the most challenging research stage, and nearly one-fifth of respondents unable to independently evaluate visual outputs, delegating instead to sighted colleagues -- not because they lacked analytic expertise, but because the tools were inaccessible~\citep{khan_i_2026}. Participants described how this forced redistribution limited their professional development and undermined the autonomy their training warranted. They also voiced distrust of specialized research tools, citing accessibility regressions, functional tools breaking after updates, and companies' performative responses to accessibility complaints~\citep{khan_i_2026}. The problem extends beyond QDA software: \citet{kumarUncoveringNewAccessibility2024} documented declining accessibility in scholarly PDFs over the past decade, and \citet{borgerMakingScienceAccessible2024} situated these barriers within systemic patterns of exclusion in research practice. \citet{keilersDataVisualizationAccessibility2023} surveyed BLV adults across multiple mediums and found that data visualization inaccessibility is pervasive, and that users want to interpret data independently.

Several themes are established through prior work: qualitative research tools are inaccessible to BLV users; this inaccessibility forces delegation that undermines epistemic autonomy; and BLV researchers distrust the tools available to them because past promises have been broken. What is limited from both the accessible visualization literature and the qualitative visualization literature are concrete steps to make qualitative data visualizations accessible, or empirical investigation of how BLV users interact with them when access is provided.
\section{Design Procedures and Goals}
\label{sec:design_procedures_and_goals}

We outline our design approach to QUARTZ, including our positionality and its impact throughout the design process, our leverage of multiple theoretical frameworks to ground our work, and the resulting design goals (DGs). 

\subsection{Positionality and Reflexivity}
\label{subsec:positionality}

As BLV researchers ourselves, we bring lived experience to both the design and evaluation of QUARTZ. The first author has low vision, and the second author is blind, a combination that shaped the system's multimodal architecture through embedded co-design (Section~\ref{subsec:design_procedures}), where R2's daily use of screen readers and refreshable Braille displays surfaced interaction barriers that heuristic evaluation alone would not have caught~\cite{mankoffDisabilityStudiesSource2010, sharifShouldSayDisabled2022}. During the evaluation, the first author's experience navigating accessibility workarounds facilitated rapport with participants and sensitivity to the troubleshooting moments that constitute some of our most consequential findings. However, our proximity to the problem also introduces potential bias: we are motivated to see QUARTZ succeed, which could lead us to underweight negative feedback or over-interpret partial task completions as successes. We mitigated this through structured data collection (task completion criteria defined a priori, facilitator interventions logged as data rather than dismissed as incidental) and by organizing findings around barriers and limitations with the same rigor as positive outcomes. We also acknowledge that our team cannot represent the full diversity of BLV experience: participants using assistive technology configurations different from our own surfaced barriers we had not anticipated, reinforcing the necessity of the RITE methodology over designer-driven testing.

\subsection{Design Procedures}
\label{subsec:design_procedures}

The design of QUARTZ was guided by three complementary frameworks. We adopted the \textit{interdependence} framework~\cite{bennettInterdependenceFrameAssistive2018} as our foundational framework, treating disability not as an individual deficit but as a relational and collective resource that shapes the design process. Interdependence has been productively employed in prior visualization work~\cite{seoMAIDRMakingStatistical2024, sharifVoxLensMakingOnline2022}, and is well-suited to our context: qualitative data analysis and visualization are inherently interpretative activities, and the perspective of a researcher who navigates such data non-visually is constitutive of what accessible qualitative analysis can and should be. 

We drew on \textit{ability-based design}~\cite{wobbrockAbilityBasedDesignConcept2011} to orient technical decisions around what users can do rather than compensating for what they cannot. In practice, this meant designing interaction and feedback mechanisms that leverage the strengths of screen reader navigation, keyboard input, and auditory cues rather than attempting to replicate visual affordances through non-visual proxies. This distinction was particularly important for qualitative visualizations, which, unlike bar charts or scatter plots, lack established non-visual interaction paradigms that designers can reference or extend~\cite{seoMAIDRMakingStatistical2024, sharifVoxLensMakingOnline2022, zongUmweltAccessibleStructured2024}. We employed \textit{participatory co-design} practices~\cite{ladner_disability_2023} throughout the development process, centering the lived experiences of BLV researchers as a primary driver of design decisions rather than treating accessibility as a post-hoc evaluation criterion. 

Our design team consisted of researchers with complementary expertise and lived experiences. The first author (R1) is a low-vision researcher with expertise in human-computer interaction and accessibility. R1 served as the primary system developer and study facilitator. The last author (R2) is a blind researcher and educator with over a decade of experience in HCI and accessibility research, and expertise in several assistive technologies. R2 served as the principal investigator, contributing domain expertise in accessible data representation, qualitative methodology, and the lived experience of conducting research as a blind scholar. R2's dual role as both co-designer and prospective end user reflects the interdependence framework's emphasis on disability experience as a design resource, rather than a subject of study~\cite{bennettPromiseEmpathyDesign2019}.

The development of QUARTZ proceeded throughout continuous, embedded co-design rather than discrete iteration phases. R1 and R2 engaged in ongoing design dialogue throughout the development process, with R2 testing prototype iterations using NVDA~\cite{nvda2025, michaelcurranNVDA202533User} on Windows, and providing real-time feedback on interaction patterns, navigation logic, information hierarchy, and auditory cues. This continuous approach was a deliberate methodological choice. Since there are limited prior accessible implementations of qualitative data visualizations, we relied on prior works' recommendations for creating accessible and interpretable qualitative visualizations~\cite{nelsonConversationTimeNew2022, pokornyNetworkAnalysisVisualization2018}. Each implemented feature generated new design questions that were surfaced through R2's direct use and resolved through collaborative discussion before subsequent features were developed. 

This embedded co-design process was generative in two respects. First, it refined the design goals outlined in Section~\ref{sec:design_goals}, which were initially derived from prior work on accessible visualization~\cite{seoMAIDRMakingStatistical2024, lundgardAccessibleVisualizationNatural2022, sharifVoxLensMakingOnline2022, blancoOlliExtensibleVisualization2022, thompsonChartReaderAccessible2023}, qualitative research methodology~\cite{milesQualitativeDataAnalysis2014, hendersonVisualizingQualitativeData2013}, and documented barriers in BLV researchers' workflows~\cite{khan_i_2026, milesQualitativeDataAnalysis2014, aishwaryaPerformingQualitativeData2022}. Repeated cycles of implementation, testing, and reflection sharpened these goals. Second, it established baseline interaction patterns and feedback mechanisms for four qualitative visualization types that were subsequently tested and refined with external BLV participants through the RITE-based evaluation described in Section~\ref{sec:evaluation}.

\subsection{Design Goals}
\label{sec:design_goals}


  

Through our co-design process, we identified four design goals that guided the development of QUARTZ. These goals address the full qualitative visualization workflow, from selecting a visualization type to refining it, to maintaining awareness of its state during exploration. They are grounded in evidence from prior work on accessible visualization, qualitative research methodology, and empirical findings on BLV researchers' workflows~\cite{khan_i_2026, nelsonConversationTimeNew2022, milesQualitativeDataAnalysis2014}.

\begin{itemize}
    \item[DG1:] \textbf{The system should empower BLV users to select visualization types through transparent recommendation scaffolding that explains tradeoffs based on data characteristics and analytical goals, enabling informed decision-making without visual preview.} Qualitative data visualizations are inherently subjective and varied in their representational approaches~\cite{nelsonConversationTimeNew2022, pokornyNetworkAnalysisVisualization2018}. BLV researchers need scaffolding to navigate these choices, as prior work demonstrates that visualization selection is challenging even for sighted users working with qualitative data~\cite{wilkinsonGrammarGraphics2012}. Moreover, BLV researchers report distrust in specialized research tools due to frequent accessibility regressions~\cite{khan_i_2026}, necessitating clear, reliable guidance from the outset of the creation process.

    \item[DG2:] \textbf{The system should enable iterative refinement through non-visual assessment, addressing both analytical validity (accuracy, clarity, complexity) and presentation quality (layout, encoding effectiveness), providing objective metrics for independent quality assurance.} Prior work shows nearly one-fifth of BLV researchers are unable to independently perform literature review or evaluate visual outputs, instead delegating tasks to sighted colleagues~\cite{khan_i_2026}. This delegation, while pragmatic, creates barriers to autonomous data exploration, delays workflows, and limits professional development opportunities. BLV researchers report that task redistribution based on accessibility constraints rather than expertise prevents them from developing diverse skills expected for career advancement.

    \item[DG3:] \textbf{The system should provide real-time complexity feedback (element count, encoding dimensions, category depth) grounded in accessibility research, recommending simplification when thresholds are exceeded while explaining rationale and allowing users to proceed with informed consent.}  Prior work on accessible visualizations for BLV users recommends limiting the number of data elements, categories, and visual encodings~\citep{elavskyHowAccessibleMy2022, kimAccessibleVisualizationDesign2021, lundgardAccessibleVisualizationNatural2022}. This is particularly critical for qualitative visualizations, which can become cognitively overwhelming when too many subjective dimensions are encoded simultaneously. Data analysis and visualization represent the most challenging research stage for BLV researchers (mean accessibility rating of 2.58/5)~\cite{khan_i_2026}, requiring systems that proactively prevent users from creating visualizations that exceed cognitive processing limits.

    \item[DG4:] \textbf{The system should maintain visualization state awareness through complementary modalities: structural descriptions via screen reader, content summaries via synthesized speech, and change notifications via audio cues—enabling incremental mental model construction.} Research shows BLV users benefit from incremental feedback during creation tasks~\cite{sharifVoxLensMakingOnline2022} and need alternative interaction paradigms beyond visual-first interfaces to understand the structure and content of their visualizations as they build them~\cite{seoMAIDRMakingStatistical2024, blancoOlliExtensibleVisualization2022}. BLV researchers express frustration with tools that fail basic accessibility requirements~\cite{khan_i_2026}, with companies offering only performative responses to accessibility concerns. Systems must provide continuous, reliable feedback to build trust and enable informed decision-making throughout the creation process.
\end{itemize}

These design goals describe the full system vision for QUARTZ as an accessible qualitative analysis and visualization environment that supports both analysis and exploration. The evaluation in this study (Section~\ref{sec:evaluation}) focuses on a foundational subset of this vision: whether BLV stakeholders can effectively navigate, comprehend, and interpret qualitative data visualizations produced by QUARTZ across four visualization types. This scoping is deliberate; refinement capabilities described in DG1-DG3 depend on users first being able to understand the visualizations under construction. If the underlying representations are not navigable and interpretable, authoring features built on top of them cannot be meaningfully assessed. Our evaluation, therefore, tests the perceptual and interaction foundations upon which QUARTZ's exploration workflow rests, surfacing barriers and informing iterative refinements through the RITE method~\cite{RapidIterativeTest2005}.

\section{System Implementation}
\label{sec:system_implementation}

\subsection{Overview}
\label{subsec:system_overview}

\begin{figure*}
    \centering
    \includegraphics[width=\linewidth]{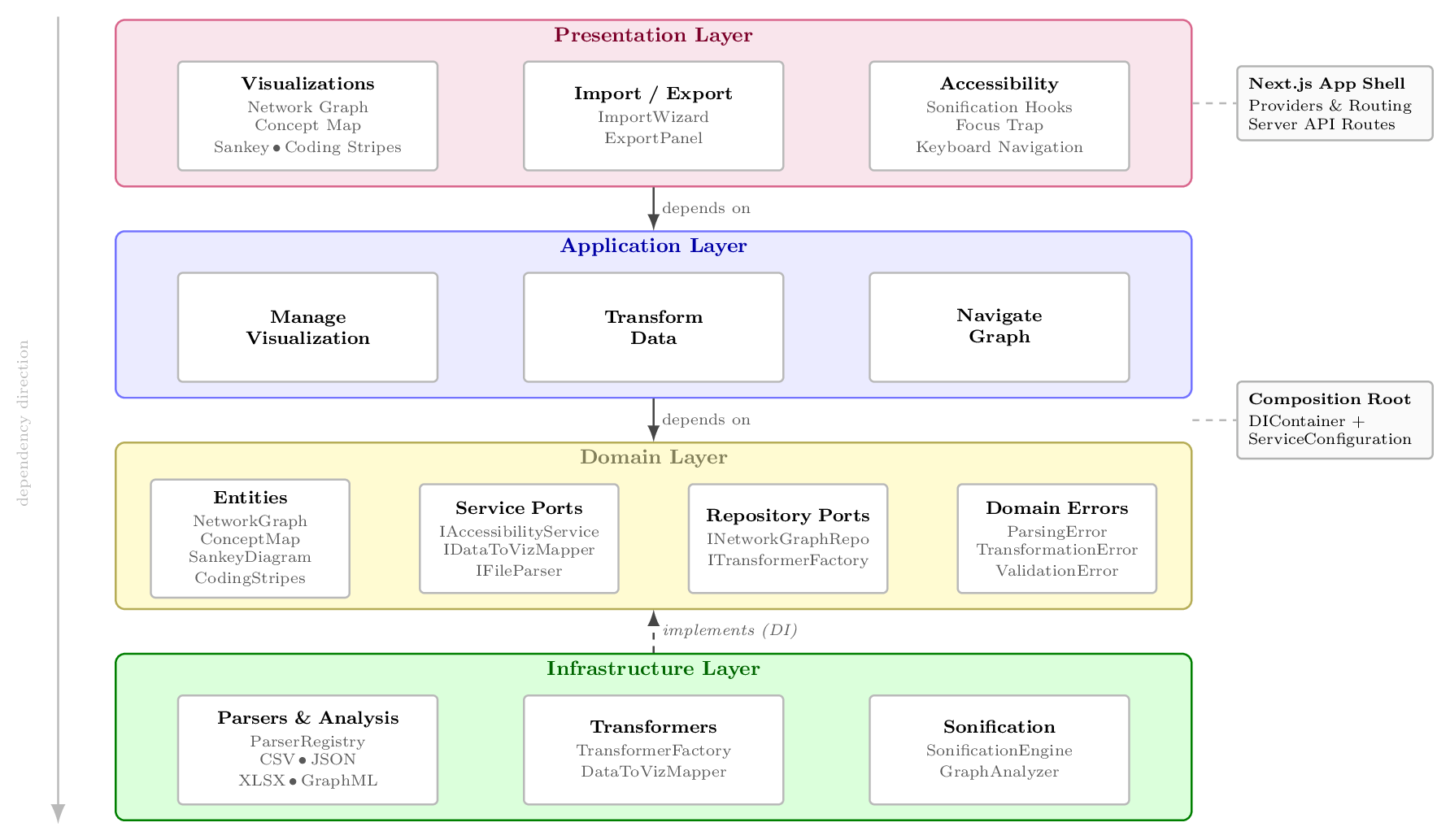}
    \caption{QUARTZ's system architecture.}
    \label{fig:system-arch-diagram}
    \Description{A four-layer system architecture diagram for QUARTZ, drawn as a vertical stack of colored horizontal bands with dependency arrows flowing top to bottom on the left side, labeled "dependency direction." The top band is the Presentation Layer (purple), containing three component groups: Visualizations (Network Graph, Concept Map, Sankey, Coding Stripes), Import/Export (ImportWizard, ExportPanel), and Accessibility (Sonification Hooks, Focus Trap, Keyboard Navigation). A downward arrow labeled "depends on" connects it to the second band, the Application Layer (blue), which contains three use-case modules: Manage Visualization, Transform Data, and Navigate Graph. Another "depends on" arrow leads to the third band, the Domain Layer (yellow), which contains four groups: Entities (NetworkGraph, ConceptMap, SankeyDiagram, CodingStripes), Service Ports (IAccessibilityService, IDataToVizMapper, IFileParser), Repository Ports (INetworkGraphRepo, ITransformerFactory), and Domain Errors (ParsingError, TransformationError, ValidationError). An upward dashed arrow labeled "implements (DI)" connects the bottom band, the Infrastructure Layer (green), which contains three concrete adapter groups: Parsers and Analysis (ParserRegistry, CSV, JSON, XLSX, GraphML), Transformers (TransformerFactory, DataToVizMapper), and Sonification (SonificationEngine, GraphAnalyzer). Two side annotations on the right identify cross-cutting concerns: the Next.js App Shell (Providers and Routing, Server API Routes) attaches to the Presentation Layer, and the Composition Root (DIContainer and ServiceConfiguration) attaches between the Application and Domain Layers. The diagram follows hexagonal architecture conventions: the Domain Layer defines abstract ports that the Infrastructure Layer implements through dependency injection, and the Presentation and Application Layers depend inward toward the Domain.}

\end{figure*}

QUARTZ (Qualitative Understanding via Accessible Representation and Visualization) is a web-based system that provides accessible, multimodal representations of qualitative data visualizations for BLV researchers. The system is built with a TypeScript stack centered around Next.js~\cite{guillermorauchNextjsVercelReact} and React~\cite{jordanwalkeReact} for the application. Supporting libraries include D3.js~\cite{mikebostockD3ObservableJavaScript} and Mermaid~\cite{sveidqvistMermaidGenerateDiagrams2014} for visualization rendering, Tone.js~\cite{mannInteractiveMusicTonejs2014}, and Web Speech for audio output. QUARTZ supports four visualization types, each grounded in an established
analytic tradition: concept maps, formalized in Novak's concept-mapping
methodology~\citep{novakTheoryUnderlyingConcept2006}; network graphs, the representational basis of qualitative
network analysis~\citep{decuypereVisualNetworkAnalysis2020, pokornyNetworkAnalysisVisualization2018}; Sankey diagrams, adapted from the flow and
stream representations used to trace how themes surface and recede over a
process~\citep{nelsonConversationTimeNew2022}; and coding stripes, drawn from the coding-display conventions
of QDA software documented by Miles, Huberman, and Saldaña~\citep{milesQualitativeDataAnalysis2014}. Each visualization is rendered with synchronized modalities, including screen reader-accessible semantic HTML with ARIA live announcements, keyboard-first structural and spatial navigation, interactive sonification, textual summary panels, and metadata descriptions. QUARTZ enables BLV users to navigate, comprehend, and interpret qualitative data without visual access. Figure~\ref{fig:system-arch-diagram} presents an overview of QUARTZ's system architecture.

A core architectural decision was to adopt a unified JSON schema that standardizes the representation of qualitative data across all four visualization types. This schema supports overlapping codes and hierarchical structures that are standard qualitative research practice (\textit{DG2}, \textit{DG3}). QUARTZ currently ingests structured, already-coded data through this schema, a tabular representation of codes, themes, segments, and relationships accepted as CSV or JSON. The schema is tool-agnostic and mirrors the export formats of major QDA packages; direct import of NVivo, ATLAS.ti, MAXQDA, and Dedoose exports, which already encode the relationships QUARTZ requires, is the nearest step on our roadmap. Unstructured sources, such as raw transcripts in word processors, require a coding pass before visualization; supporting them, first through already-coded documents and comment threads and ultimately through an in-tool coding pass from raw text, is a priority direction rather than out of scope.    

\subsection{Architecture}
\label{subsec:system_architecture}

Figure~\ref{fig:four_images_main} (Section~\ref{subsec:rw_qual_viz}) illustrates each visualization type as rendered in QUARTZ.

\subsubsection{Visualization Selection and Recommendation Scaffolding (\textit{DG1})}
\label{subsubsec:dg1}

QUARTZ addresses \textit{DG1} by making the visualization selection process transparent and challengeable by BLV researchers. When data are imported, a detection layer infers the best-fitting visualization type using structural heuristics: datasets with average connection density exceeding two per node are routed to the network graph representation; sparser datasets suggest concept maps or Sankey diagrams, depending on whether hierarchical or flow-oriented structures dominate. A complementary column-pattern analyzer scores hierarchical, network, and flow structures in tabular imports using keyword weights and column-level statistics, selecting the highest-scoring match above a minimum confidence threshold. These heuristics provide an initial recommendation, but QUARTZ treats the recommendation as a starting point rather than a final decision. The system's role is to scaffold the researcher's judgment, not replace it. To support that judgment, QUARTZ provides a \textit{Why this type?} panel, a focus-trapped, keyboard-navigable dialog that explains the reasoning behind the current visualization type. The panel presents five categories of information: (1)~the data structure detected in the imported dataset, (2)~explanatory reasons why the current visualization type matches the data, (3)~a mapping table showing how data fields correspond to visual and non-visual representation elements, (4)~per-type complexity constraints with status indicators (\textit{within range}, \textit{near limit}, or \textit{exceeds recommended range}), and (5)~alternative visualization types rated as \textit{good fit}, \textit{possible}, or \textit{poor fit} with accompanying reasoning. Complexity thresholds are tailored to each visualization type; for instance, concept maps flag node counts outside 3--100 or hierarchy depths beyond 2--6 levels, while network graphs evaluate edge-to-node ratio and Sankey diagrams assess whether stage counts fall within 3--7. Upon opening, the panel announces a summary via an \texttt{aria-live} region: the visualization type, how many complexity constraints are satisfied, and how many alternatives are rated as a good fit, enabling BLV researchers to quickly gauge whether the current representation warrants deeper exploration or whether an alternative should be considered. The detection layer is entirely rule-based; no generative model is involved at any stage. The complete keyword weights, per-structure scoring contributions, structural bonuses, confidence gate, and per-type readiness point allocations are documented in Appendix~\ref{appendix:implementation-details} (Tables~\ref{tab:detection-layer} and~\ref{tab:readiness-allocations}). These values are deliberate calibrations for our setting rather than empirically optimized constants, following the tradition of rule-based visualization recommendation~\citep{mackinlay_automating_1986}.

\subsubsection{Non-Visual Iteration (\textit{DG2})}
\label{subsubsec:dg2}

QUARTZ supports iterative refinement of qualitative visualizations without visual access through keyboard-driven editing affordances that provide non-visual feedback at each step. In concept maps, researchers can edit node descriptions via an accessible \texttt{textarea} and create cross-links through a dedicated link mode: pressing \texttt{L} activates the mode, \texttt{Enter} selects source and target nodes in sequence, and a dialog collects the relationship type and label before the connection is confirmed. Network graphs extend this pattern with node repositioning via \texttt{Shift+Arrow} keys and a history stack that supports undo and redo across node creation, deletion, and link modification. Each editing action produces non-visual feedback that closes the edit-confirm loop: earcons confirm cross-link stages (source selected, target selected, link created, or error), and \texttt{aria-live} regions announce outcomes such as the newly created relationship and connected nodes. This ensures that BLV researchers receive confirmation of structural changes through the same modalities they use to navigate, without needing to inspect the updated rendering visually.

These GUI-level affordances are underpinned by a bidirectional synchronization layer that converts between typed domain entities, the structured models used for rendering, sonification, and screen reader output, and a Mermaid text representation. This architecture serves two purposes: it maintains a single source of truth across interaction modes and enables an alternative text-based editing path via a Monaco-based code editor with syntax validation and live preview. When the GUI and text representations diverge, QUARTZ surfaces a conflict resolution dialog offering \textit{Keep GUI Changes} and \textit{Keep Code Changes} with an optional diff view, rather than silently discarding either version. Our user study (Section~\ref{sec:evaluation}) evaluated the GUI editing path; the text-based editing mode remains available as an advanced capability for researchers who prefer direct code manipulation.

\subsubsection{Accessible Complexity Feedback (\textit{DG3})}
\label{subsubsec:dg3}

QUARTZ provides structured, quantified feedback on visualization complexity
and structural readiness through an Evaluation Panel that presents an
accessible quality assessment for each visualization type. The panel reports
a readiness score on a 1-5 scale (labeled "Publication Readiness" in the
interface), calculated as a weighted ratio of passed quality checks: checks classified as errors carry triple weight, warnings carry double, and informational checks carry single weight. The resulting score maps to discrete labels: Excellent~(5), Good~(4), Acceptable~(3), Needs Work~(2), or Poor~(1), and an export readiness threshold of 3 determines whether the
visualization is flagged as ready for export. Upon opening, the panel announces a summary via an \texttt{aria-live} region: "Evaluation panel opened for Concept Map. Publication readiness score: 4 out of 5, Good. 6 of 7 quality checks passed. Ready for export."

Quality checks are tailored to each visualization type and evaluate structural completeness, reasonable complexity, and labeling adequacy. Concept map checks verify that all nodes are connected, node counts fall within 3-100, hierarchy depth remains between 2-6 levels, a root concept exists, and all edges carry labels. Network graph checks additionally evaluate edge-to-node ratio bounds and, when layout metrics are available, node occlusion, edge crossing percentage, and edge tunneling against established readability thresholds. Sankey diagram checks assess flow balance at intermediate nodes (inflow-outflow divergence within 10\%) and the presence of clear source and sink nodes. Coding stripes checks verify that code coverage exceeds 10\% of total text and that code counts fall within 2-50. Each check specifies a severity level and a plain-language description that explains both the issue and its recommended resolution, making the feedback actionable rather than purely diagnostic.

The panel also presents a generated structure summary that contextualizes the quality checks with a prose description of the visualization's composition. For example, "This concept map contains 14 concepts and 19 relationships. The hierarchy has a maximum depth of 3 levels. All concepts are connected. Root concept: Remote Work Analysis." Keyboard navigation within the panel is supported through jump buttons (Readiness, Export, Summary, Statistics, Checks) that scroll to and focus the corresponding section heading, enabling screen reader users to move directly to the information most relevant to their current concern rather than reading the panel linearly. We scope what these checks do and do not verify. They evaluate structural completeness, complexity bounds, and labeling adequacy, properties that can be computed from the visualization's data model without visual inspection. They do not evaluate visual-design dimensions such as label positioning, color palette suitability, or aesthetic layout quality, and the readiness score should not be read as a comprehensive assessment of presentation quality. Where a check has a corresponding WCAG success criterion, the implementation targets it; for example, the Sankey contrast adjustment made during the study (refer to Appendix~\ref{subsec:rite-appendix}) brought stage rendering to WCAG AA. Extending the check suite toward the visual-design dimensions that sighted reviewers assess is future work directly motivated by the epistemic verification barrier in Section~\ref{subsubsec:epistemic-barriers}.

\subsubsection{Multimodal State Awareness (\textit{DG4})}
\label{subsubsec:dg4}

\begin{figure*}[t]
    \centering
    \includegraphics[width=\linewidth]{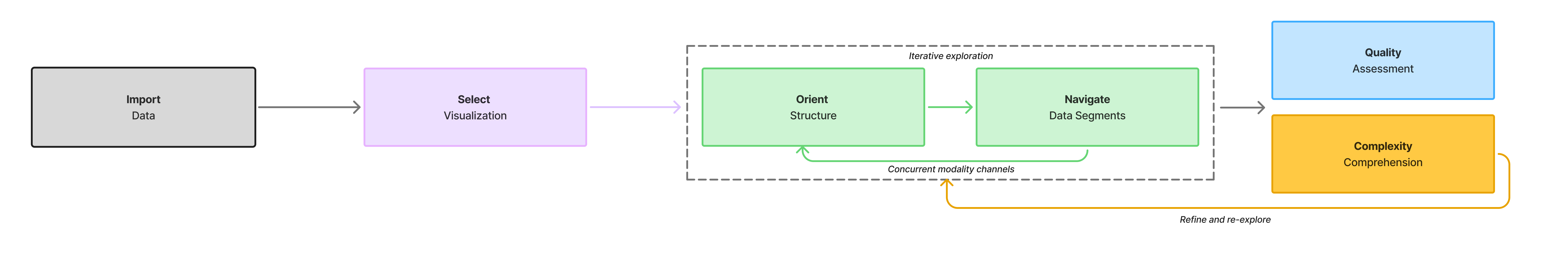}
    \caption{A user's workflow through QUARTZ.}
    \label{fig:user_workflow}
    \Description{A horizontal workflow diagram showing a user's path through QUARTZ as five connected stages, read left to right. Stage 1, "Import Data," is a dark gray box. An arrow leads to Stage 2, "Select Visualization," shown as a purple box. Another arrow leads into a dashed rectangle labeled "Iterative exploration" that groups two green boxes: "Orient Structure" and "Navigate Data Segments." An arrow connects Orient to Navigate, and a return arrow underneath loops from Navigate back to Orient, labeled "Concurrent modality channels," indicating that users move back and forth between orienting to structure and navigating segments rather than proceeding linearly. An arrow exits the dashed group and splits toward two parallel endpoints on the right: a light blue box labeled "Quality Assessment" on top and an orange box labeled "Complexity Comprehension" on the bottom. A curved feedback arrow, labeled "Refine and re-explore," loops from the Complexity Comprehension box all the way back to the Orient Structure box, indicating that comprehension outcomes can send the user back to the iterative exploration loop. Color coding distinguishes the stage types: gray for data input, purple for selection, green for the iterative exploration pair, and blue and orange for the parallel assessment outcomes.}
\end{figure*}

QUARTZ renders each visualization with four synchronized non-visual modalities: semantic HTML with ARIA live announcements, interactive sonification, keyboard-first structural navigation, and textual summary panels. These modalities are coordinated so that a single navigation action updates the screen reader, plays a sonification tone, and shifts keyboard focus simultaneously; the user controls which channels are active and at what level of detail.

\paragraph{Screen reader layer.} Each visualization uses ARIA roles appropriate to its data structure. Concept maps employ \texttt{tree} and \texttt{treeitem} roles with \texttt{aria-level}, \texttt{aria-expanded}, and \texttt{aria-selected} attributes; the \texttt{A}~key toggles between a detailed label (concept name, description, level, child count, and enumerated cross-links with relationship types) and a concise label (name, level, and aggregate connection counts). Network graphs use \texttt{role="application"} on the SVG container with an assertive live region that announces the current node's label, cluster membership, connection count, and strongest connection upon each navigation step. Sankey diagrams annotate each focusable node with its stage position (e.g., "Stage 2 of 5, node 3 of 4") alongside flow values. Coding stripes announce segment text, applied codes, and positional context within the document.

\paragraph{Sonification.}
QUARTZ maps structural properties of each visualization type to auditory parameters (refer to Table~\ref{tab:sonification_mappings}). Concept maps use a Web Audio sine oscillator with pitch derived from hierarchy depth: a base frequency of 261.63~Hz (C4) is raised by a half-octave per tree level, producing progressively higher tones for deeper concepts; leaf nodes are distinguished by a separate triangle-wave earcon at 130~Hz. Network graphs employ a Tone.js synthesis engine that maps normalized degree centrality to pitch (220-880~Hz), betweenness centrality to volume, and Louvain community membership to stereo pan, spreading clusters across the stereo field; a cluster chord overview (\texttt{O}~key) plays up to four notes per community sorted by centrality for a rapid structural scan. Sankey diagrams map flow magnitude to pitch (200-800~Hz), normalized across all links in the diagram. Coding stripes map code identity to fixed pitch values and segment length to tone duration, so longer coded passages produce proportionally longer tones. Shared earcons provide audio confirmation for discrete actions including mode transitions, link creation, and navigation boundaries.

\begin{table}[t]
\centering
\caption{Default sonification parameter mappings across QUARTZ's four visualization types.
Users can remap the pitch, volume, and pan channels for network graphs via the encoding
configuration panel.}
\label{tab:sonification_mappings}
\begin{tabular}{p{1cm}lllp{1.5cm}}
\toprule
\textbf{Viz} & \textbf{Pitch} & \textbf{Volume} & \textbf{Pan} & \textbf{Duration} \\
\midrule
Concept Map    & Tree depth         & Fixed       & Mono      & Fixed (300\,ms)  \\
Network Graph  & Degree centrality  & Betweenness & Community & Fixed (250\,ms)  \\
Sankey Diagram & Flow magnitude     & Fixed       & Mono      & Fixed (300\,ms)  \\
Coding Stripes & Code identity      & Fixed       & Mono      & Segment length   \\
\bottomrule
\end{tabular}
\end{table}

\paragraph{Keyboard navigation.} Each visualization implements a navigation pattern suited to its data structure. Concept maps follow the ARIA treeview pattern: \texttt{Up/Down Arrow} moves between concepts in linearized tree order, \texttt{Right Arrow} expands a node or moves into its subtree, \texttt{Left Arrow} collapses or returns to the parent, and \texttt{Tab} opens a cross-links quick view for navigating non-hierarchical connections. Network graphs use a browse-then-commit model: \texttt{Left/Right Arrow} cycles through a node's connections, \texttt{Enter} follows the highlighted connection, \texttt{Backspace} returns to the previous node, and \texttt{Up/Down Arrow} switches between outgoing and incoming connection lists; a dedicated \texttt{I}~key provides a detailed structural description including cluster membership, bridge status, and ranked connections. Sankey diagrams use a stage-grid model where \texttt{Left/Right Arrow} moves between flow stages and \texttt{Up/Down Arrow} moves between nodes within a stage. Coding stripes use segment-level roving focus, with arrow keys advancing through coded text segments and announcements identifying the applied codes and position within the document.

\paragraph{Modality coordination and user control.} Navigation handlers are structured so that a single keystroke triggers updates across all active modalities; for example, cycling to a new connection in the network graph calls the sonification engine and sets the live region announcement text within the same event handler. Users can independently toggle sonification (\texttt{S}), mute narration (\texttt{M}), and adjust screen reader verbosity (\texttt{A}), foregrounding the modality most useful for their current task. QUARTZ applies sensible defaults for each visualization type (Table~\ref{tab:sonification_mappings}) and additionally exposes an encoding configuration panel that allows remapping of which structural property drives pitch, volume, and pan for network graphs, as well as an audio settings panel for adjusting pitch range and master volume. All preferences persist across sessions via local storage.

\subsection{User Workflow}
\label{subsec:user_workflow}


Figure~\ref{fig:user_workflow} depicts a typical journey through QUARTZ for concept maps. On load, QUARTZ announces a structural overview ("Concept map. 14 nodes, 19 connections, 3 thematic clusters") (\textit{DG4}). The researcher presses Tab to enter the visualization, hears the first concept with its level and child count alongside a guidance prompt listing the available keys, and traverses the hierarchy with arrow keys, with each stop announcing the concept's connections and relationship types. Pressing Space plays a sonification cue whose pitch encodes hierarchy depth. When ready, the researcher activates the Evaluation button for a readiness summary (\textit{DG3}): the readiness score, passed checks, and a prose structure summary. Throughout, the researcher controls which clusters to visit, which modalities to attend to, and what level of detail to request; the A key toggles between detailed and concise screen reader modes, eliminating the need for visual access or sighted assistance.
\section{Evaluation}
\label{sec:evaluation}

We evaluated QUARTZ through a user study employing the Rapid Iterative Testing and Evaluation (RITE) method~\cite{RapidIterativeTest2005} with 8 BLV participants. Below, we describe our participants, apparatus, procedure, and analytic approach.

\subsection{Study Design}
\label{subsec:study_design}

We chose the RITE method~\cite{RapidIterativeTest2005} for two reasons specific to our design context. First, there are limited accessible implementations of qualitative data visualizations~\cite{blancoOlliExtensibleVisualization2022, aishwaryaPerformingQualitativeData2022}, which means there are limited established interaction patterns or design guidelines we could have relied upon to anticipate the barriers users would encounter. We primarily relied on prior recommendations for qualitative visualization design~\cite{wilkinsonGrammarGraphics2012, oanceaQualitativeNetworkAnalysis2017, decuypereVisualNetworkAnalysis2020, nelsonConversationTimeNew2022}, but limited implementations exist. RITE's iterative structure, in which the research team reviews session data after each participant and implements targeted modifications before the next session, allowed us to surface and address accessibility barriers in real time rather than documenting them passively across an entire study. Second, the diversity of BLV users' assistive technology configurations (screen readers, operating systems, navigation strategies) means that barriers surfaced by one participant may not arise for another, and vice versa. RITE enabled us to respond to this diversity as it emerged.

Unlike traditional usability testing, in which the system remains fixed throughout the study, and modifications are deferred to a subsequent design cycle, RITE treats the evaluation itself as a design intervention. Changes made between participants are documented as data and analyzed as part of the findings. This means that later participants used an improved version of the system compared to earlier participants, as we discuss in Section~\ref{sec:findings}.

The evaluation focused on the exploration and comprehension of qualitative data visualizations, testing the foundational interaction layer on which QUARTZ's authoring capabilities (\textit{DG1-DG3}) rest. \textit{DG4} (multimodal state awareness) was directly evaluated, as it operates during both exploration and creation (refer to Section~\ref{sec:design_goals} for further discussion).

\subsection{Participants}
\label{subsec:participants}

We designed a screening survey to gather high-level insights with qualitative data workflows, including closed-ended Likert scale questions to understand confidence and familiarity, and open-ended questions to allow for further elaboration. We then requested and received approval for our study from our organization's Institutional Review Board (IRB). Subsequently, we contacted several community partners for permission to distribute the survey, including University of Washington's DO-IT Center~\footnote{https://doit.uw.edu/} electronic mailing lists (AccessComputing and AccessSTEM), the National Federation of the Blind (NFB)~\footnote{https://www.nfb.org}, and the National Research and Training Center on Blindness and Low Vision (NRTC)~\footnote{https://www.blind.msstate.edu/} at Mississippi State University. To be eligible for the study, participants had to meet the following Inclusion criteria:

\begin{itemize}
    \item 18 years of age or older
    \item Conversational proficiency in English 
    \item Identify as being blind or having low-vision (corrected visual acuity of 20/70 or less)
    \item Be actively engaged in research as an individual contributor OR have previously participated in research as an individual contributor 
\end{itemize}

Upon completion of the scheduled session, participants were compensated with a USD 50 Amazon e-gift card. Table~\ref{tab:participants} summarizes participant demographics, assistive technology configurations, and relevant experience. 

\begin{table*}[t]
  \centering
  \caption{Participant demographics and assistive technology configurations.}
  \label{tab:participants}
  \begin{tabular}{lllll}
    \toprule
    P\# & Vision Status & Assistive Technology & OS & Research Exp. \\
    \midrule
    P1  & Legally blind & JAWS & Windows & 3  \\
  P2  & Low vision & ZoomText and OS Magnification & macOS & 12 \\
  P3  & Legally blind & JAWS & Windows & 4 \\
  P4  & Legally blind & NVDA & Windows & 5 \\
  P5  & Legally blind & ZDSR & Windows & 5 \\
  P6  & Legally blind & NVDA & Windows & 3 \\
  P7  & Low vision & Screen and OS Magnification & Windows & 15 \\
    P8  & Low vision & Screen and OS Magnification & Windows & 20 \\
    \bottomrule
  \end{tabular}
\end{table*}

Documenting assistive technology configurations is important because, as prior work has shown~\cite{seoMAIDRMakingStatistical2024}, the behavior of accessible web applications can vary significantly across screen readers and operating systems. Differences in how JAWS, NVDA, and VoiceOver handle ARIA roles, live regions, and focus management directly affect the user experience and can surface barriers that are specific to particular AT configurations.

\subsection{Apparatus}
\label{subsec:apparatus}

Participants interacted with the QUARTZ system in their preferred desktop-based browser. The first author conducted all sessions remotely via Zoom. This modality was intentional so participants could use their own assistive technology configurations, which increased ecological validity but introduced variability.

The study used three qualitative visualization stimuli across the four supported types. Each visualization was constructed from sample coded interview data created for the study. The stimuli were designed to represent realistic qualitative data at moderate complexity, large enough to exercise every navigation and sonification path, but small enough to remain tractable within a single task window. Specifically, the concept map stimulus ("Remote Work Challenges") was a four-level hierarchy of 23 themes connected by 22 parent-child links; the network graph stimulus ("Code Co-occurrences") comprised 7 codes connected by 12 weighted co-occurrence edges (weights 1--8); and the Sankey stimulus ("Job Search Experiences") traced 18 nodes across 5 sequential stages with 20 flows. The coding stripes stimulus reused the same Job Search interview that seeded the Sankey diagram: a single 875-word transcript annotated with 26 coded segments drawn from an 8-code codebook spanning strategy, emotion, barriers, coping, and outcomes. Stimuli were author-generated and structured to follow widely used qualitative analysis conventions: hierarchical theming~\cite{novakTheoryUnderlyingConcept2006}, code co-occurrence networks, and process-stage flows~\cite{milesQualitativeDataAnalysis2014, lundgardAccessibleVisualizationNatural2022}, so that the data read as plausible coded fieldwork while staying free of any participant-identifying content.

\subsection{Tasks}
\label{subsec:tasks}

\begin{figure}[ht]
    \centering
    \begin{subfigure}[b]{0.48\textwidth}
        \centering
        \includegraphics[width=\linewidth]{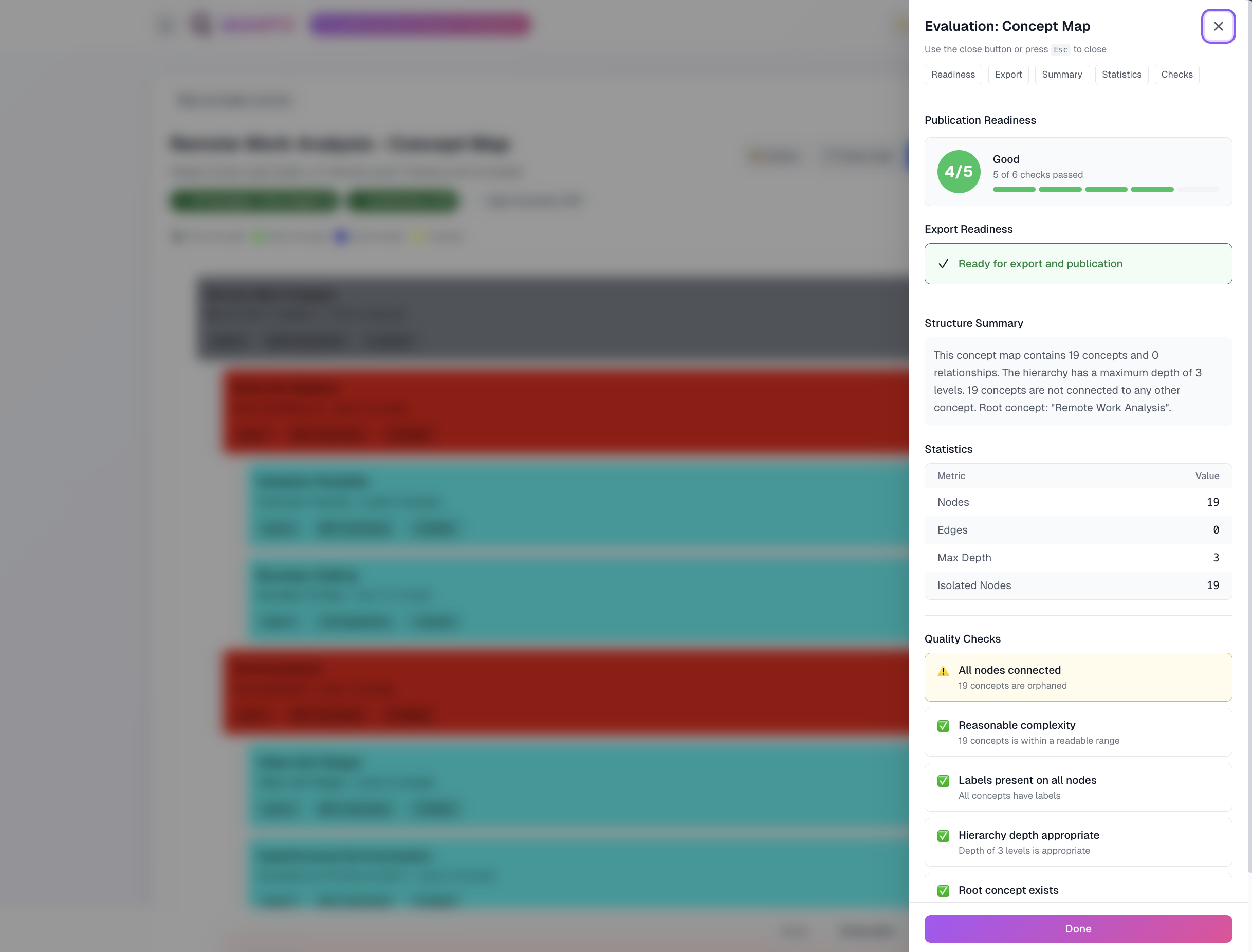}
        \caption{QUARTZ's visualization readiness evaluation modal.}
        \label{fig:eval-modal}
    \end{subfigure}\hfill
    \begin{subfigure}[b]{0.48\textwidth}
        \centering
        \includegraphics[width=\linewidth]{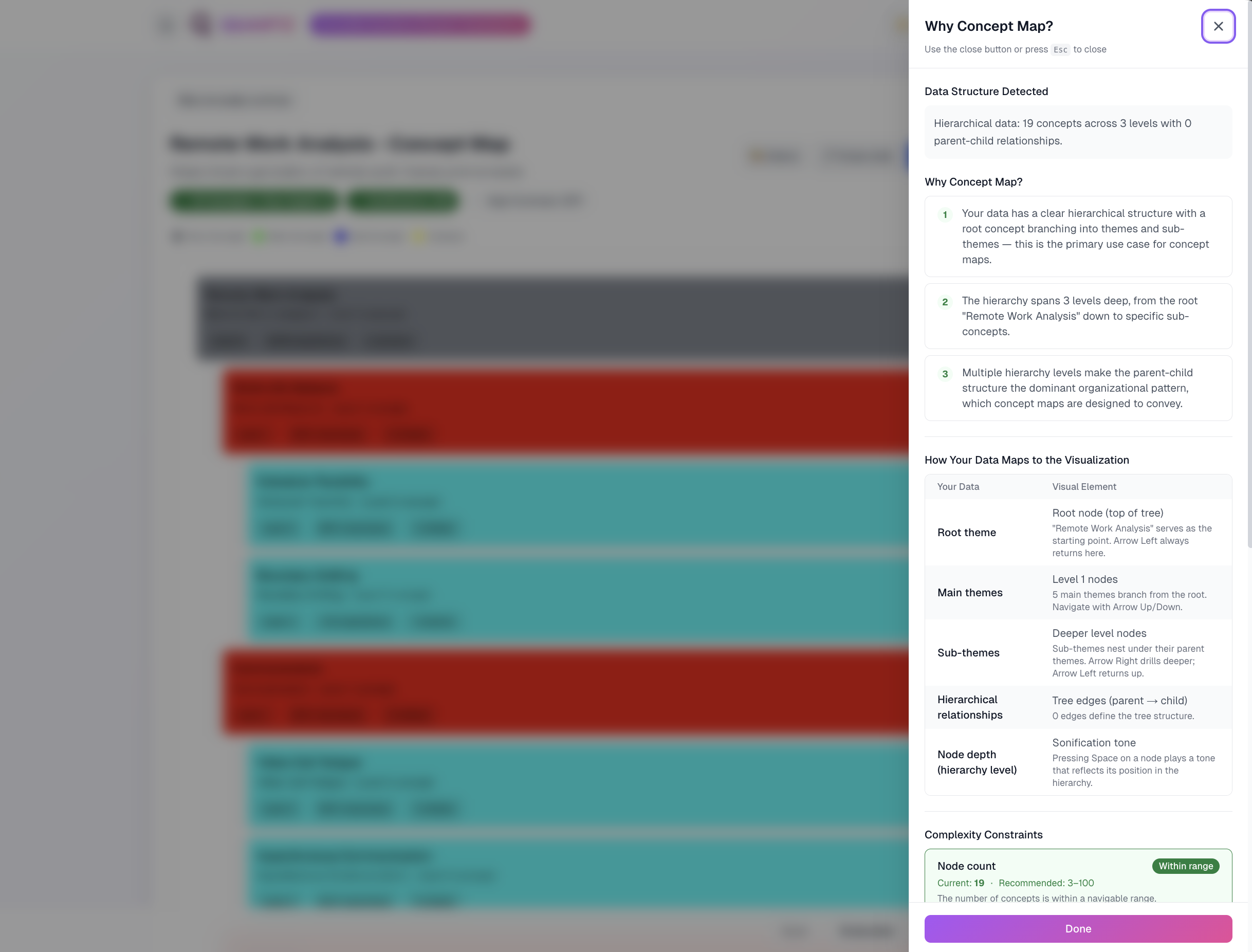}
        \caption{QUARTZ's decision-making modals.}
        \label{fig:why-this-type-modal}
    \end{subfigure}
    \caption{Screenshots of modals from QUARTZ that inform presentation readiness and rationale for visualization selection.}
    \label{fig:presentation-modals}
    \Description{Two screenshots of modal panels from QUARTZ, shown side by side. The left panel (a), titled "Evaluation: Concept Map," reports a Publication Readiness score of 4 out of 5 (labeled "Good, 5 of 6 checks passed") and an Export Readiness status of "Ready for export and publication." It includes a Structure Summary stating the concept map contains 19 concepts and 0 relationships across a hierarchy of 3 levels, with the root concept "Remote Work Analysis." A Statistics table reports 19 nodes, 0 edges, max depth 3, and 19 isolated nodes. A Quality Checks section lists one warning ("All nodes connected: 19 concepts are orphaned") and four passed checks: reasonable complexity, labels present on all nodes, hierarchy depth appropriate, and root concept exists. The right panel (b), titled "Why Concept Map?," shows a Data Structure Detected summary ("Hierarchical data: 19 concepts across 3 levels with 0 parent-child relationships") followed by three numbered reasons the system recommends a concept map: the data has a clear hierarchical structure, the hierarchy spans 3 levels, and multiple hierarchy levels make the parent-child structure the dominant organizational pattern. A "How Your Data Maps to the Visualization" table maps data elements to visual elements: root theme to root node, main themes to level 1 nodes, sub-themes to deeper level nodes, hierarchical relationships to tree edges, and node depth to sonification tone. A Complexity Constraints section indicates that node count (19) is within the recommended range of 3 to 100. Both panels include a close button at the top right and a Done button at the bottom. The main page behind each modal is blurred.}
\end{figure}

Participants completed 12 tasks distributed across the four visualization types (3 tasks per type). Tasks were designed to assess three levels of comprehension:

\begin{itemize}
  \item \textbf{Orientation tasks} assessed whether participants could identify the overall structure of a visualization: the number of elements, their organization, and the general topology (RQ1).
  \item \textbf{Navigation and retrieval tasks} assessed whether participants could locate specific elements, follow connections between data elements, and retrieve targeted information from the visualization (RQ1, RQ2).
  \item \textbf{Interpretation tasks} assessed whether participants could derive meaning from the visualization: identifying patterns, making comparisons, or forming analytic judgments about the qualitative data represented (RQ1, RQ3, RQ4).
\end{itemize}

Task design was informed by the visualization task taxonomies used in prior accessible visualization research~\cite{seoMAIDRMakingStatistical2024, kimAnsweringQuestionsCharts2020}, adapted for qualitative visualization types. For example, where MAIDR's tasks asked participants to "find the maximum value" in a bar chart (a lookup task with a single correct answer), our orientation tasks asked participants to describe the overall structure of a concept map---a task that requires synthesizing relational information rather than retrieving a discrete data point.

\subsection{Procedure}
\label{subsec:procedure}


Each session lasted approximately 120 minutes and followed a consistent structure.

\paragraph{Setup and orientation (\textasciitilde 10 minutes).} The facilitator (R1) confirmed the participant's assistive technology configuration, ensured the QUARTZ system loaded correctly in their browser, and provided a brief orientation to the system's interface and keyboard shortcuts. Participants were encouraged to explore the system freely during this period and to ask questions.

\paragraph{Task completion (\textasciitilde 90 minutes).} Participants completed the 12 tasks in fixed order, blocked by visualization type. For each task, the facilitator read the task prompt aloud and asked the participant to think aloud as they worked. Participants were given a 5-minute time limit to complete each task. Following each task, participants were asked to rate their confidence in completing the task on a 7-point Likert scale, with 1 indicating "not confident at all", up to 7 indicating "extremely confident".

\paragraph{Post-study interview (\textasciitilde 10 minutes).} Following task completion, the facilitator conducted a semi-structured interview exploring the participant's overall experience with QUARTZ, their strategies for navigating different visualization types, their perceptions of the system's usefulness for qualitative research, and suggestions for improvement. The facilitator administered the Accessibility Usability Scale (AUS) survey~\cite{alwarpillaiAccessibleUsabilityScale} to obtain quantitative insights on participants' experiences with QUARTZ.     

\paragraph{RITE iterations.} Consistent with the RITE methodology, the research team reviewed session data after each participant, including task completion outcomes, think-aloud observations, facilitator notes on accessibility barriers encountered, and instances of facilitator intervention. When a barrier was identified that could be addressed through a targeted system modification, the change was implemented and documented before the next session. All modifications were recorded in a change log (refer to Appendix~\ref{subsec:rite-appendix}) that captures the session in which the issue was identified, a description of the barrier, the modification made, and the design rationale.

\subsection{Data Collection and Analysis}
\label{subsec:design_and_analysis}

We collected multiple forms of data across sessions:

\begin{itemize}
  \item \textbf{Task performance data:} task completion status (success, partial, failure), time on task derived from session recording timestamps, and facilitator intervention frequency per task.
  \item \textbf{Think-aloud data:} audio recordings and transcripts of participants'  verbalizations during task completion.
  \item \textbf{Facilitator observation notes:} real-time notes documenting accessibility barriers encountered, instances of facilitator intervention, and the nature of each intervention.
  \item \textbf{Post-study interview transcripts:} audio recordings and transcripts of semi-structured interviews.
  \item \textbf{RITE change log:} documented modifications to QUARTZ between sessions, including the barrier that motivated each change, the modification made, and the outcome observed in subsequent sessions.
  \item \textbf{Post-task questionnaire:} closed-ended AUS survey~\cite{alwarpillaiAccessibleUsabilityScale} data and open-ended responses to questions about the most valuable aspect of the experience, the most challenging aspect, likelihood of adoption in the participant's own research workflow, and desired features or improvements.
\end{itemize}

We analyzed qualitative data (think-aloud transcripts, interview transcripts, facilitator notes, open-ended questionnaire responses) using inductive thematic analysis~\cite{braunUsingThematicAnalysis2006}. Following each session, the first author reviewed the recording and transcript, annotated facilitator interventions with their type and triggering barrier, and identified emergent usability and accessibility issues. Issues that could be resolved without altering the study's core interaction design were addressed between sessions as part of the RITE methodology; the change, its rationale, and the session in which it was introduced were recorded in the change log (refer to Appendix~\ref{subsec:rite-appendix}). After all sessions were completed, the first author conducted initial open coding across the full corpus of think-aloud and interview transcripts. Codes were iteratively organized into themes spanning four areas corresponding to our research questions: (1)~visualization selection and complexity scaffolding, capturing how participants used recommendation guidance and complexity feedback to orient their exploration (RQ1); (2)~independent creation and workflow comparison, capturing participants' experiences creating visualizations relative to their current approaches involving general-purpose tools or sighted assistance (RQ2); (3)~multimodal feedback and navigation, capturing how participants used screen reader descriptions, synthesized speech summaries, and audio cues to maintain autonomy and control during exploration and creation (RQ3); and (4)~non-visual quality assessment, capturing how participants evaluated visualization quality through QUARTZ's assessment mechanisms and their perceptions of these mechanisms for independent research practice (RQ4). The last author participated in regular check-ins throughout the analysis process, reviewing candidate themes, challenging interpretations, and discussing alternative groupings until both authors reached consensus on the final thematic structure. Findings are organized by research question in Section~\ref{sec:findings}.
 
Facilitator interventions were categorized into five types to enable systematic comparison of intervention frequency and nature across early versus later sessions: (1)~\textit{navigational guidance}, in which the facilitator directed the participant to a specific UI region or element (e.g., "navigate to the home button at the top left of the screen"); (2)~\textit{shortcut instruction}, in which the facilitator introduced or reminded the participant of a keyboard shortcut (e.g., "you can press L to enter link mode"); (3)~\textit{error recovery}, in which the facilitator helped the participant return to a known state after an unintended action or loss of focus; (4)~\textit{task clarification}, in which the facilitator restated or elaborated on the current task objective; and (5)~\textit{conceptual explanation}, in which the facilitator explained what a visualization element or relationship represented in the context of the data. We distinguish between interventions that reflect system-level accessibility gaps (navigational guidance, error recovery) and those inherent to the think-aloud protocol for a novel system (shortcut instruction, task clarification, conceptual explanation); this distinction informs our interpretation of intervention reduction across RITE iterations and is reported alongside accessibility barriers across all four RQs.
 
Quantitative data (task completion, time on task, AUS scores) were analyzed descriptively. Given the small sample size and the RITE methodology's inherent introduction of system changes across participants, we do not draw inferential statistical conclusions but instead use quantitative measures to contextualize qualitative findings and to characterize the trajectory of improvement across sessions. Task completion outcomes are reported across RQ1-RQ3 where relevant to each question's scope; AUS scores are reported under RQ4 (Section~\ref{subsec:findings_rq4}); and intervention frequency and RITE trajectory are reported as a cross-cutting theme that spans all four research questions.
 
We also report facilitator intervention; the facilitator provided navigational guidance when participants encountered accessibility barriers that prevented task progress. The frequency and nature of these interventions are documented as data: they represent the accessibility barriers that QUARTZ's initial design did not adequately address, and their reduction across sessions represents the effect of RITE iterations.
\section{Findings}
\label{sec:findings}

We present findings organized by our four research questions (RQs), followed by a cross-cutting analysis of accessibility barriers and the RITE iterations that addressed them. Barriers span all four RQs, so we consolidate them in Section~\ref{subsec:findings_barriers} rather than fragmenting them across individual questions.

 \subsection{RQ1: Visualization Selection and Complexity Scaffolding}
\label{subsec:findings_rq1}

To address RQ1, we designed a visualization wizard to guide BLV users toward qualitative visualization types that match their data while constraining complexity for non-visual comprehension.

\subsubsection{Selection Scaffolding}
\label{subsubsec:selection-scaffolding}

Participants responded positively to the recommendation scaffolding (\textit{DG1}), though several requested additional depth. Its most valued property was \textit{transparency}, explaining \textit{why} a visualization type was recommended based on data characteristics, not simply asserting \textit{what} was recommended.

P2 drew a direct contrast with prior tools:

\begin{quote}
    "\textit{I've in the past tried to generate diagrams or visualizations, and gotten either an incomplete or not very helpful visualization, or just a pop-up window saying 'warning, there isn't enough data', an error message that's incomplete. Like, 'this won't work,' but I don't understand why. [QUARTZ] [gives] a pretty comprehensive description as to why something may or may not be [suitable].}" --P2
\end{quote}

\noindent This contrast, opaque failure versus transparent recommendation, highlights the distrust BLV researchers hold toward tools that make unexplained decisions, and also echoes prior work in exploring research workflows~\cite{khan_i_2026}. It also reinforces prior accounts of accessibility shortcomings of specialized research tools~\cite{aishwaryaPerformingQualitativeData2022}.

P7, self-described as "pretty inexperienced with those visualizations," valued the pedagogical orientation: 

\begin{quote}
    "\textit{The system made an effort to introduce me to these different visualizations and to provide guidance in thinking about which one could be most suitable for my data set.}" --P7
\end{quote}

\noindent P7 suggested it could go further with "\textit{even a sample of what each of these types of visualizations look like}", a preview capability QUARTZ does not currently offer.

However, not all participants found the scaffolding sufficiently detailed. P4 called the descriptions "a bit vague" and wanted "\textit{a more detailed preview, so [they] know exactly what [they are] getting.}" P5 pushed further, requesting conversational interactivity: 

\begin{quote}
    "\textit{I wish there was a small chat box. I can type some questions, why [did] you recommend this? And what's the foundation for you to make this decision?}" --P5
\end{quote}

\noindent P5 also articulated a broader tension: 

\begin{quote}
    "\textit{It made 99\% of the decision for me, and \textbf{there's not enough space for me to do the personalization}. That made me a little bit uncomfortable.}" --P5 
\end{quote}

\noindent This tension between the guidance BLV users need and the agency they desire is a design challenge static scaffolding alone cannot resolve. It requires real-time, interactive feedback to clarify system decisions so BLV users feel greater confidence in visualization selection. This is critical given the more flexible nature of qualitative visualizations compared to their quantitative counterparts.  

P6 engaged critically with the readiness scores, noting that "one was at 35\%, the network graph" and questioning the rubric: "I\textit{ wish [QUARTZ] gave a more specific rubric to the percentage.}" Quantitative readiness indicators were noticed and used, but not fully trusted without explanatory context.

\subsection{RQ2: Independent Exploration and Creation vs. Current Workflows}
\label{subsec:findings_rq2}

To answer RQ2, we assessed to what extent QUARTZ enables BLV users to
independently explore and perform structured creation tasks on
qualitative data visualizations compared to their current workflows.

\subsubsection{Current Workflow Practices}
\label{subsubsec:current-workflows}

Participants' current approaches ranged from full delegation to complete avoidance. P4 described the delegation pattern documented in prior work~\cite{khan_i_2026}:

\begin{quote}
    "\textit{A lot of times when creating these diagrams, I have a set of colleagues who create them, and then they will describe it to me, so I would understand kind of what's going on, but I wouldn't be able to directly interact with it as much as this one.}" --P4
\end{quote}

\noindent P8 also reported having "no experience creating [qualitative] visualizations with any other software". From this, we can glean how the inaccessibility of existing tools has not merely forced delegation but prevented engagement entirely. 

\subsubsection{Creation Experience with QUARTZ}
\label{subsubsec:creation-experience}

The creation experience was generally positive, particularly during data import. P1 described the process as "easier" because "\textit{it does it for [them]}" rather than requiring manual inspection. P8 also compared it favorably: "\textit{It's a lot better than Qualtrics, I'll give you that.}" P7 found the import process "\textit{pretty comparable to other import processes.}"

However, the creation workflow's most valued property was \textit{direct interaction with output}. P4 captured this shift, stating that although their current workflow involves colleagues creating and describing visualizations, QUARTZ allowed them to "\textit{just read it [themselves] and go through the data [themselves].}" P6 articulated the same principle:

\begin{quote}
    "\textit{The most valuable aspect is \textbf{being able to interact with the data, rather than just being given a description of what the data is about}.}" --P6
\end{quote}

P5 valued the \textit{standardization} of each visualization ("\textit{no matter how much skill [I] have, [I] can at least generate the standard visualization for [my] presentation}") but wanted more customization, highlighting a recurring tension between automated decisions and researcher control that also surfaced in RQ1. We note the scope of these creation experiences: all creation tasks (cross-links, annotations, code application) operated on pre-structured study data rather than participants' own research projects. We therefore read them as evidence of structured-task capability within an exploration workflow, not as a demonstration of end-to-end authoring from raw qualitative data, which we identify as future work in Section~\ref{subsec:limitations_and_future_directions}.

\subsubsection{Adoption Readiness}
\label{subsubsec:adoption-readiness}

All participants expressed some likelihood of adoption, though several attached conditions. These conditions were multi-faceted in nature. One factor was \textit{AT integration}, as observed when P4 made JAWS compatibility the gating factor: "\textit{if the virtual cursor issue is resolved, meaning that with the virtual cursor on, I can use the keyboard shortcuts.}" Another condition was \textit{cross-platform reliability}, as highlighted by P5: "\textit{I [want] it to work well with different laptops, or different browser and screen reader.}" P3 echoed reliability concerns, calling the system "a really welcome development" given that accessible qualitative visualization offerings are "so lacking,", but noted it "didn't seem as seamless as [they] hoped."

These responses echo the distrust documented by \citet{khan_i_2026}: participants valued what QUARTZ offers but assessed it against a history of tools that promised accessibility and failed to deliver.

\subsection{RQ3: Multimodal Feedback for Autonomy and Control}
\label{subsec:findings_rq3}

When approaching RQ3, we assessed how BLV users experience QUARTZ's multimodal feedback mechanisms to support autonomy and control through structured tasks that could be completed through multiple modalities.

\subsubsection{Navigation Strategies Across Visualization Types}
\label{subsubsec:navigation-strategies}

Participants approached the four visualization types with distinct strategies shaped by each type's structure.

\paragraph{Concept Maps} The hierarchical tree structure was the most navigable type across all visualizations. Participants used arrow keys to traverse hierarchy levels and expand child nodes, and relied on screen reader feedback to learn about each concept's properties. P3 found level indicators particularly orienting: "\textit{What's helpful is when it says 'level 1', 'level 2', when it tells you the level.}" The tree structure provided a natural reading order that mapped well to screen reader traversal and expedited learnability of the map as a whole.

\paragraph{Network Graphs} In contrast, network graphs were consistently the most challenging. P6 identified \textit{spatial configuration} as one of the core problems when attempting to piece together the graph's structure: 

\begin{quote}
    "\textit{When we were going through the network graphs, it's just telling me 'frustration connected to adaptation, resilience.' Being able to make the network graph section more interactive, for the other diagrams, I'm able to press up and down and explore the diagram...[I hope] the network graph area can improve so I can explore the diagram and see what is there for myself}" --P6     
\end{quote}

The absence of hierarchical structure meant the navigation paradigm successful in concept maps did not transfer. This was echoed by several other participants (P3, P5), and also played a significant role in lower task completion rate compared to the other visualization types.  

\paragraph{Sankey Diagrams} Sankey diagrams produced the most consistent navigation experience across participants. The flow context of the arrow keys, left and right to traverse stages, up and down to select among flows within a stage, mapped directly onto participants' mental models of a job search pipeline. As stated by P3: 

\begin{quote}
    "\textit{This layout of stage 1 through 6, with the left-right arrow, go up and down for each cut from that stage, and then having the value read out loud, I think that makes sense to me.}" --P3
    
\end{quote}

P4 and P6 experienced no navigation interventions during their Sankey tasks, and P6 specifically noted that the boundary announcements ("already at the first stage," "already at the last stage") helped them track position. Sankey diagrams also elicited the strongest positive response to sonification (further discussed in Section~\ref{subsubsec:modality-prefs}), which together with the flow-based arrow keys made this type the most accessible of the four we tested.

\paragraph{Coding Stripes} Coding stripes were the most sequentially structured type, producing fewer orientation barriers but surfacing a distinct category of text-selection barriers. P1 reported that interacting with coded text "\textit{helped [them] connect dots better}" compared to reading the transcript directly. For participants who completed the code application task successfully, the workflow was compared favorably against delegating coding to sighted colleagues. The barriers that did arise concentrated on two functionalities: entering text-selection mode reliably, a shortcut-discovery barrier in early sessions, and receiving feedback about what text had been selected, an AT-rendering barrier that persisted across screen readers. These were most acute for participants using JAWS virtual cursor (P4) or ZDSR (P5), where \texttt{Shift-Arrow} key input was either intercepted or not announced.

Figure~\ref{fig:task_completion} summarizes task completion outcomes across the four visualization types, showing a clear divide between pre- and post-RITE participants (P1--P4 vs.~P5--P8) for coding stripes tasks, reflecting the text-selection barriers resolved mid-study.

\begin{figure}[h]
  \centering
  \footnotesize
  \setlength{\tabcolsep}{4pt}
  \renewcommand{\arraystretch}{1.15}
  \begin{tabular}{@{}l|cccc!{\color{okDivider}\vrule width 1pt}cccc@{}}
    \toprule
     & P1 & P2 & P3 & P4 & P5 & P6 & P7 & P8 \\
    \midrule
    \multicolumn{9}{@{}l}{\textit{Concept Map}} \\
    T1 Orientation    & \cS & \cS & \cP & \cP & \cS & \cS & \cS & \cS \\
    T2 Navigation     & \cP & \cS & \cS & \cS & \cP & \cS & \cS & \cS \\
    T3 Interpretation & \cS & \cS & \cS & \cS & \cS & \cS & \cS & \cS \\
    \addlinespace
    \multicolumn{9}{@{}l}{\textit{Network Graph}} \\
    T4 Orientation    & \cP & \cS & \cF & \cF & \cP & \cF & \cS & \cP \\
    T5 Navigation     & \cP & \cS & \cP & \cF & \cP & \cF & \cS & \cP \\
    T6 Interpretation & \cP & \cS & \cP & \cP & \cP & \cF & \cS & \cS \\
    \addlinespace
    \multicolumn{9}{@{}l}{\textit{Sankey Diagram}} \\
    T7 Orientation    & \cS & \cS & \cP & \cS & \cS & \cS & \cS & \cS \\
    T8 Navigation     & \cS & \cS & \cP & \cS & \cS & \cS & \cS & \cS \\
    T9 Interpretation & \cS & \cS & \cP & \cS & \cP & \cP & \cS & \cP \\
    \addlinespace
    \multicolumn{9}{@{}l}{\textit{Coding Stripes}} \\
    T10 Orientation    & \cS & \cS & \cP & \cS & \cP & \cP & \cS & \cP \\
    T11 Navigation     & \cP & \cS & \cP & \cP & \cS & \cP & \cS & \cP \\
    T12 Interpretation & \cS & \cS & \cS & \cS & \cS & \cS & \cS & \cS \\
    \bottomrule
  \end{tabular}

  \vspace{0.4em}
  \begin{tabular}{@{}l@{\hspace{1em}}l@{\hspace{1em}}l@{}}
    \cellcolor{okSuccess}\textcolor{white}{\ \ding{51}\ } Success &
    \cellcolor{okPartial}\textcolor{black}{\ $\boldsymbol{\sim}$\ } Partial &
    \cellcolor{okFailure}\textcolor{white}{\ \ding{55}\ } Failure \\
  \end{tabular}

  \caption{Task completion outcomes across the 12 study tasks (rows, grouped by visualization type) and 8 participants (columns). The heavier vertical divider separates sessions before (P1-P4) and after (P5-P8) the major RITE modifications reported in Section~\ref{subsec:findings_barriers}.}
  \label{fig:task_completion}
  \Description{A 12-by-8 grid of task-by-participant outcomes. Rows are grouped by visualization type. Concept Map rows (T1, T2, T3) are all success for every participant. Network Graph rows show: T4 success for P1 through P7 and partial for P8; T5 partial for P1 and success for P2 through P8; T6 partial for P1, P3, and P6, success for the rest. Sankey Diagram rows (T7, T8, T9) are all success. Coding Stripes rows show: T10 success for P1 through P7 and partial for P8; T11 partial for P1 through P4 and success for P5 through P8; T12 partial for P3 and P4, success for the rest. A heavier vertical rule separates P1 through P4 from P5 through P8 to mark the pre- and post-RITE split.}
\end{figure}

\subsubsection{Modality Preferences}
\label{subsubsec:modality-prefs}

Participants exhibited diverse modality preferences, consistent with prior work~\cite{seoMAIDRMakingStatistical2024, blancoOlliExtensibleVisualization2022}, but with patterns specific to qualitative visualization structure.

\paragraph{Sonification} P2 identified sound-based feedback as the most valuable aspect of the system: 

\begin{quote}
    "\textit{The sound-based feedback and tonality, I navigate both tech and other environments with a lot of audio integration, that's my preference. Having an option that's built into the site, that can be turned on and off.}" --P2

\end{quote}

P6 provided the most detailed account, specifically for Sankey diagrams:

\begin{quote}
    "\textit{Being able to hear a sound played, and if the sound is higher, you know that the flow is wider. If the sound is lower pitched, you know that it is thinner, and so you can easily go through the diagram and get a picture of what it might look like. With other systems, sometimes they just give you a number, and then you have to think, okay, so this is 20, this is 10, so this part should be wider. You're just kind of imagining that, but with this one, \textbf{you are able to tell right away the shape and what the diagram would look like}.}" --P6
\end{quote}

\noindent This captures a qualitative-specific contribution: sonification conveying \textit{shape}, a spatial property, rather than numerical magnitude. P6 extended this to quality confidence: 

\begin{quote}
    "\textit{If I were to print this out or something, I have more confidence that it's going to look the way that I wanted it to look.}" --P6
    
\end{quote}

\paragraph{Screen Reader Descriptions} Structural descriptions via screen reader served as the primary orientation modality for the majority of participants. For instance, P3 valued hierarchy metadata in concept maps. For network graphs, however, P6's critique that the system was "\textit{just telling [them] frustration connected to adaptation}" suggests that listing connections without relational context is insufficient for non-hierarchical structures, a limitation of textual descriptions that sonification partially addresses.

\paragraph{Audio Cues} Beyond Sankey sonification, non-speech audio cues served two roles across visualization types: signaling hierarchy position and confirming state transitions. In concept maps, pitch was mapped to hierarchy level: lower tones for parent nodes at level 0, higher tones for children at deeper levels. Participants who triggered the sonification during concept map traversal reported that this helped them track their position in the tree without requiring an explicit screen reader announcement. State-transition cues (entering or exiting the visualization area, activating cross-link mode) were also consistently noticed; P2's appreciation of the toggle ("\textit{can be turned on and off}") explicitly extended to these cues, not only to sonification. However, not every participant engaged with audio cues. P1 initially reported that the sonification shortcut "didn't work" and required facilitator guidance to activate it, and P3's sessions included periods where audio cues played during screen reader output, creating overlap that obscured both channels. These observations suggest that audio cues function best when paired with clear activation affordances and careful timing against concurrent speech output.

\subsubsection{Autonomy and Control}
\label{subsubsec:autonomy-and-control}

The multimodal architecture's most valued property was the shift from \textit{passive reception} to \textit{active exploration}. As per P6: 

\begin{quote}
    "\textit{The most valuable aspect is being able to interact with the data, rather than just being given a description.}" --P6
    
\end{quote}

P4's contrast between their current workflow ("\textit{colleagues create them and describe it to me}") and QUARTZ ("\textit{read it yourself and go through the data yourself}") captures the autonomy shift \textit{DG4} was designed to enable. P5 found the system "\textit{helpful for [them] to quickly switch between some links [they] built up}," supporting the iterative back-and-forth that qualitative analysis requires.

However, autonomy in navigation did not fully resolve autonomy in verification. P5, despite indicating high task confidence (7/7) for several tasks, expressed concern:

\begin{quote}
    "\textit{Because I don't have the vision, \textbf{I really worry what it's showing is really not what I really want to express. It's like AI or this tool have fully made a decision for me}, so it makes me feel less confident about my visualization.}" --P5
\end{quote}

\noindent This anxiety about whether non-visual representations accurately mirror visual output is a barrier that multimodal feedback alone cannot resolve, and motivates the quality assessment mechanisms examined in RQ4.

\subsection{RQ4: Non-Visual Quality Assessment}
\label{subsec:findings_rq4}

To investigate RQ4, we explored how BLV users assess visualization quality through non-visual evaluation mechanisms: the Visualization Readiness panel, the "Evaluate" panel, and the "Why This Type?" panel.

\subsubsection{Quality Assessment Practices}
\label{subsubsec:quality-assessment}

Participants scrutinized quality indicators rather than accepting them at face value. Beyond the rubric requests reported in Section~\ref{subsubsec:selection-scaffolding}, P8 wanted "some sort of quick guide" to understand detected limitations; indicators were noticed and used, but not self-explanatory. P5 valued standardization as a form of quality assurance: "No matter how much skill you have, you can at least generate the standard visualization." This positions quality assessment not as expert review but as a floor of competence, assurance that the output follows conventions a sighted audience would recognize. P4's observation that QUARTZ enables them to "read it [themselves] and go through the data [themselves]" suggests that independent assessment, navigating the output rather than receiving a colleague's description, was itself perceived as quality assurance, even beyond explicit metrics. These findings suggest that non-visual assessment mechanisms begin to address the gap documented by ~\citet{khan_i_2026}, where nearly one-fifth of BLV researchers cannot independently evaluate visual outputs; however, the metrics require more contextual explanation to be fully trusted.

\subsubsection{Usability Scores}
\label{subsubsec:usability-scores}

\begin{figure}
    \centering
    \includegraphics[width=\linewidth]{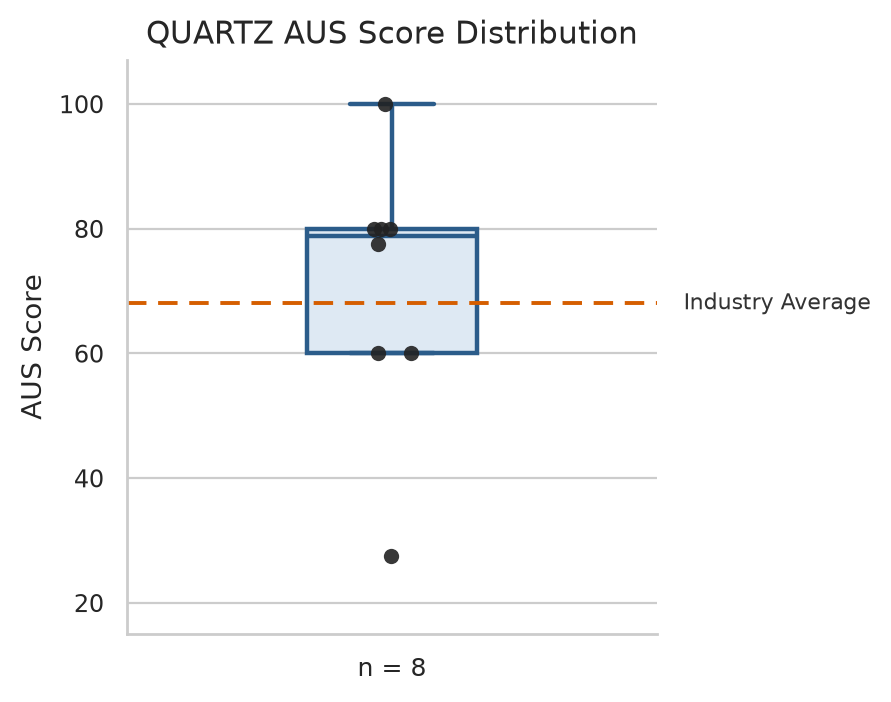}
    \caption{Participants' post-session AUS scores.}
    \label{fig:aus-plot}
    \Description{A vertical box plot of post-session AUS scores from 8 participants, on a y-axis ranging from roughly 15 to 105. The box extends from 60 at the first quartile to 80 at the third quartile, with the median line at approximately 79. The upper whisker reaches 100; there is no lower whisker because the lower quartile coincides with the lowest non-outlier score. Individual participant scores are overlaid as dots: one at 100, three at 80, one at 77.5, two at 60, and one low outlier at 27.5. A dashed horizontal line at 68, labeled Industry Average, marks the conventional benchmark; the box and seven of the eight scores sit at or above the box's lower edge near this line, while the single outlier sits well below it. The plot is titled QUARTZ AUS Score Distribution and is annotated n = 8 beneath the x-axis.}
\end{figure}



Figure~\ref{fig:aus-plot} depicts the AUS score distribution across all participants. QUARTZ achieved an above-average mean AUS score ($\bar{x} = 70.6$, $s = 21.6$). Because RITE modifications altered the system between sessions, these scores aggregate experiences across successive system versions. We therefore read them as a property of the design trajectory rather than a benchmark of the final build, and defer summative usability claims to a future fixed-system evaluation.

\subsection{Accessibility Barriers and RITE Iterations}
\label{subsec:findings_barriers}

Accessibility barriers surfaced across all four RQs. We consolidate them here as they share a common root cause: the structural, semantic, and AT-dependent properties of qualitative visualizations that existing paradigms do not address.

\subsubsection{Structural Navigation Barriers} 
\label{subsubsec:barrier-cats}

The most pervasive barriers involved \textit{focus management}. P6 described repeated episodes where "\textit{it's reading from the top, 'data visualization!', when I'm trying to go to a section, it kept on jumping back to the top.}" P1 could not independently locate the visualization area, asking, "\textit{[How] do I go there?}" and requiring verbal spatial description from the facilitator. Network graphs presented a distinct challenge. The absence of hierarchical organization meant the tree-navigation paradigm, successful in concept maps, did not apply. P6 noted that while concept maps, Sankey diagrams, and coding stripes allowed pressing "up and down" to "explore the diagram," the network graph section listed connections without supporting interactive traversal. This asymmetry confirms that qualitative visualizations cannot be treated as a single accessibility problem; each type's topology demands its own navigation model.

\subsubsection{Vocabulary Barriers} 
\label{subsubsec:vocab-barriers}

Participants unfamiliar with qualitative visualization encountered conceptual rather than technical barriers. P7 identified "\textit{understanding the common terminology, like nodes and edges, and thinking through how that applies to a dataset}" as their most challenging experience. P8 echoed this: "\textit{Not being familiar with the terminology}" was their main difficulty. These barriers do not arise in quantitative chart accessibility, where terms like bar, line, and pie are more widely understood. This suggests that accessible qualitative visualization systems must scaffold the conceptual vocabulary of data representation, a challenge \textit{DG1's} recommendation scaffolding only partially addresses.

\subsubsection{Epistemic Verification Barriers} 
\label{subsubsec:epistemic-barriers}

Separate from vocabulary barriers, some participants could navigate the visualization structure but could not verify that it expressed their analytic intent. P5, despite completing tasks with high confidence, worried: "\textit{because [they] don't have the vision, [they] really worry what it's showing is really not what [they] really want to express.}" This is not a failure of navigation or comprehension; P5 understood what the visualization contained. It is a failure of \textit{verification}: the inability to confirm that the non-visual representation successfully mirrors the visual output a sighted colleague would see. This barrier is distinct from quantitative chart accessibility, where the mapping from data to visual encoding is deterministic. In qualitative visualization, the mapping is interpretive, making verification inherently more difficult.

\subsubsection{AT Compatibility Barriers} 
\label{subsubsec:at-barriers}

The most consequential AT barrier involved JAWS's virtual cursor conflicting with QUARTZ's keyboard shortcuts. P4 reported that "\textit{sometimes it's working, sometimes it's not, it's just the main issue with the JAWS virtual cursor}," and conditioned adoption on resolution. P5 encountered screen reader failures on ZDSR: "\textit{Somehow my screen reader cannot read out very well}." P2 reported possible conflicts between their laptop's built-in narrator and QUARTZ's key commands. P3 experienced content not being read aloud during coding stripe interaction, describing the system as "\textit{not necessarily complex or hard to use, but just not fully seamless}."

These barriers emphasize that accessible qualitative visualization cannot be validated against a single AT configuration. The diversity we observed: JAWS cursor conflicts, ZDSR rendering failures, narrator interference, is consistent with prior work~\cite{seoMAIDRMakingStatistical2024, fanAccessibilityDataVisualizations2023} but particularly acute for qualitative visualizations, whose non-standard DOM structures stress-test ARIA implementations more aggressively than linear chart layouts.

\subsubsection{RITE Trajectory}
\label{subsubsec:rite-trajectory}

\begin{figure}[b]
  \centering
  \includegraphics[width=\linewidth]{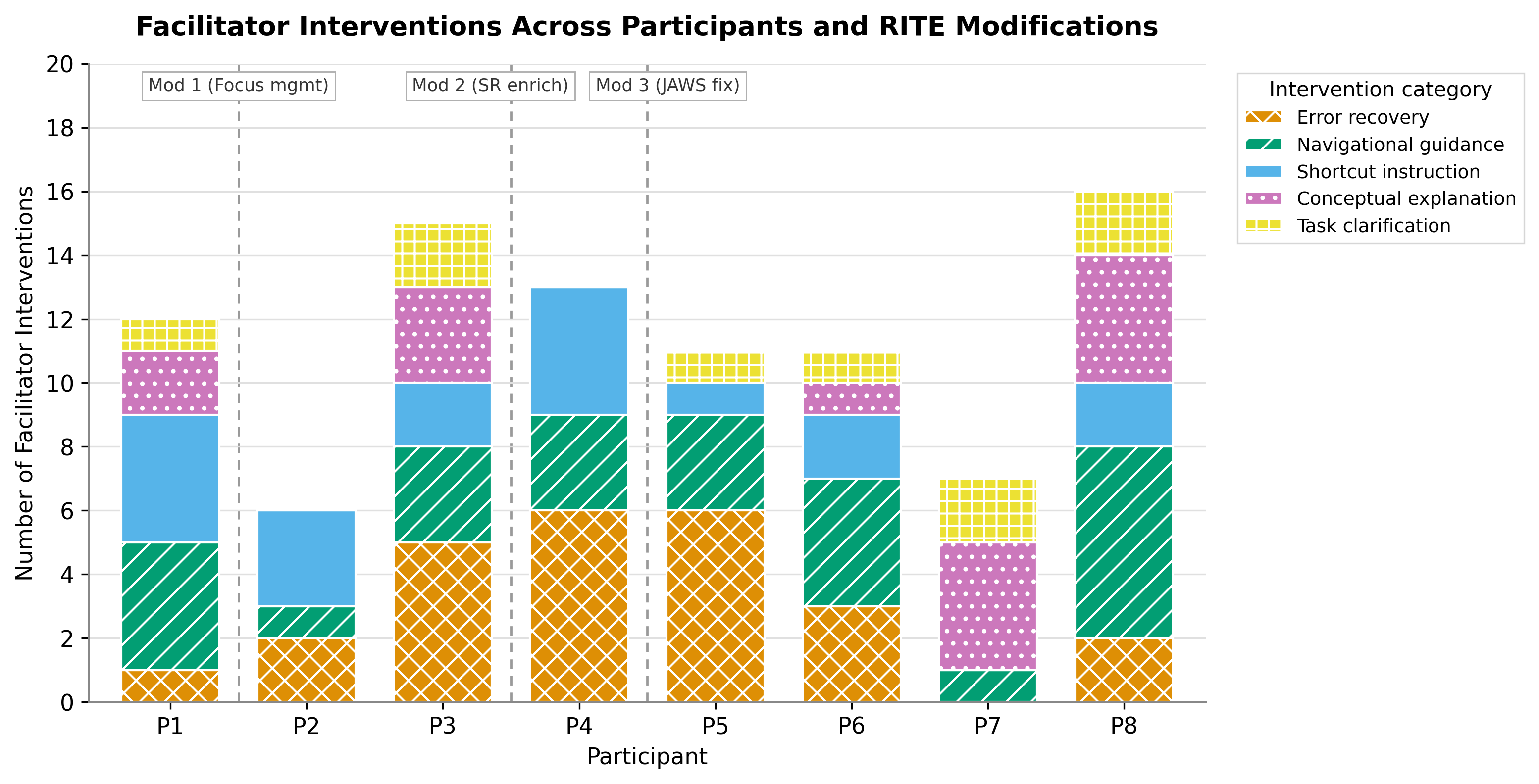}
  \caption{Facilitator intervention frequency across participants (P1-P8).}
  \label{fig:trajectory}
  \Description{Stacked bar chart titled "Facilitator Interventions Across Participants and RITE Modifications." The x-axis displays eight participants (P1–P8); the y-axis shows intervention count (0–20). Five intervention categories are stacked using a colorblind-safe palette with distinct hatch patterns for monochromatic accessibility: Error recovery (crosshatched), Navigational guidance (forward-hatched), Shortcut instruction (solid), Conceptual explanation (dotted), and Task clarification (cross-hatched). Intervention totals per participant: P1=12, P2=6, P3=15, P4=13, P5=11, P6=11, P7=7, P8=16. Dashed vertical lines mark modification boundaries: Mod 1 (Focus management) after P1, Mod 2 (Screen reader enhancement) after P3, and Mod 3 (JAWS fix) after P4. Error recovery and Navigational guidance are the most common intervention types, together accounting for 62\% of all interventions. The figure uses a light gridline background, light gray grid lines, and a legend positioned to the right.}
\end{figure}

Figure~\ref{fig:trajectory} visualizes the trajectory of improvement (refer to Appendix~\ref{subsec:rite-appendix} for detailed documentation of each modification).

Early sessions with heavy AT conflicts (P3: 15, P4: 13) required the most substantial facilitator guidance. P1's session (12 interventions) surfaced fundamental focus-management barriers; P1 could not independently navigate into the visualization area and required spatial descriptions from the facilitator. P3's session revealed screen-reader rendering failures and multiple system bugs (cross-link count bug, \texttt{R} key structure-summary-only bug, display modal not opening on \texttt{D} key) that required facilitator acknowledgment and workarounds (5 error-recovery interventions in total). P4's session was dominated by JAWS virtual cursor conflicts, which intercepted QUARTZ's keyboard shortcuts and forced facilitator-guided workarounds throughout the session (6 error-recovery interventions, the highest of any session).

P2 was an outlier among early sessions (6 interventions total), reflecting their prior experience with accessible visualization tools and strong verbal fluency; they self-recovered from several keyboard shortcut failures rather than requiring facilitator intervention. Following modifications to focus management (after P1) and enriched screen reader announcements (after P3), P5 and P6 showed similar totals (11 each), though their error-recovery counts remained high (P5: 6, P6: 3). P5's session was dominated by ZDSR compatibility issues, an AT not previously tested, where \texttt{V}, \texttt{D}, \texttt{R}, \texttt{K}, and \texttt{A} keys variously failed to register, requiring the facilitator to take remote control to position focus manually. This indicates that the RITE modifications addressed JAWS-specific conflicts but did not generalize to all screen readers. After JAWS passthrough handling was implemented following P4's session, P7 recorded the lowest total intervention count of the study (7), with zero error-recovery interventions. However, P7 used zoom and mouse as their primary interaction modality rather than keyboard shortcuts, so this reduction partly reflects their modality rather than solely the system improvements.
 
P8's session returned to the highest count (16); their interventions were not AT-conflict driven: they were weighted toward navigational guidance (6) and conceptual explanation (4). P8 surfaced vocabulary barriers ("\textit{parent/child has nothing to do with remote work}"), UI signifier confusion (buttons vs. non-clickable tiles), and zoom/pan out-of-bounds problems. None of these categories was targeted by Modifications 1-3. This pattern illustrates that RITE's trajectory is shaped by the barriers participants encounter, not only by system state: new participants can surface new barrier categories that prior modifications did not address~\cite{RapidIterativeTest2005}. P8's session also indicates that eight sessions did not saturate the barrier space. We treat the categories they surfaced, vocabulary scaffolding, UI signifier clarity, and zoom and pan bounds for low-vision interaction, as the agenda for the next RITE cycle rather than as closed findings, and we identify a larger follow-up study as future work (Section~\ref{subsec:limitations_and_future_directions}).
 
Because RITE modifications alter the system between participants, direct comparison across sessions reflects both system improvement and individual differences. We interpret these patterns as indicative of design trajectory rather than controlled experimental evidence.

\subsubsection{System Modifications}
\label{subsubsec:rite-modifications}

%

\begin{figure*}[t]
    \centering
    
    \subcaptionbox{Workspace entry and re-entry guidance (Modification~1 target).\label{fig:mod-ba-a}}[0.32\textwidth]{%
        \includegraphics[width=\linewidth]{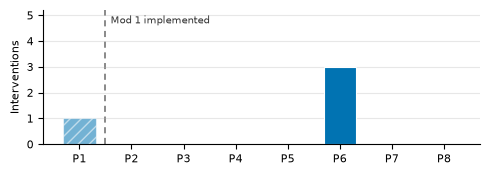}%
    }\hfill
    \subcaptionbox{Relational-meaning explanations in network graphs (Modification~2 target). P4's zero is structural: the network graph was skipped in his session.\label{fig:mod-ba-b}}[0.32\textwidth]{%
        \includegraphics[width=\linewidth]{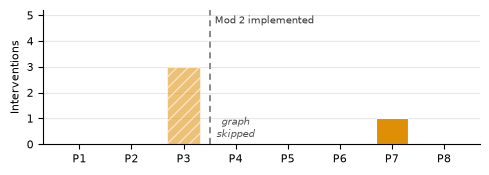}%
    }\hfill
    \subcaptionbox{Shortcut failures caused by assistive technology input interception (Modification~3 target). P5 used ZDSR, an assistive technology not previously tested; P7--P8 used zoom and mouse rather than keyboard shortcuts.\label{fig:mod-ba-c}}[0.32\textwidth]{%
        \includegraphics[width=\linewidth]{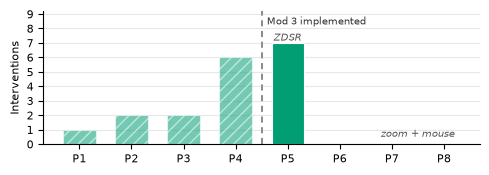}%
    }
    
    \centering
    \caption{Interventions matching each modification's target barrier, per participant. Dashed lines mark when each modification was implemented; hatched bars precede the modification and solid bars follow it.}
    \label{fig:mod_before_after}
    \Description{Three stacked bar charts, one per modification, each showing interventions per participant from P1 to P8 with a dashed line marking when the modification was implemented. Chart a, workspace entry and re-entry guidance: P1 has 1 intervention before Modification 1; afterward, P2 through P5 have zero, P6 has 3, and P7 and P8 have zero. Chart b, relational-meaning explanations in network graphs: P3 has 3 interventions before Modification 2; afterward, P7 has 1 and all others have zero, with P4's zero annotated as structural because the network graph was skipped in his session. Chart c, shortcut failures from assistive technology interception: P1 has 1, P2 has 2, P3 has 2, and P4 has 6 before Modification 3; afterward, P5 has 7 (annotated ZDSR, a screen reader not previously tested), P6 has zero, and P7 and P8 have zero (annotated zoom plus mouse, indicating they did not rely on keyboard shortcuts). All y-axes use whole numbers.}
\end{figure*}

%
%


Figure~\ref{fig:mod_before_after} isolates each modification's target barrier across sessions, complementing the aggregate trajectory in Figure~\ref{fig:trajectory}. The pattern is instructive rather than uniformly favorable. Workspace-entry guidance (Figure~\ref{fig:mod-ba-a}) disappeared immediately after Modification 1 for P2 through P5, though P6 required three re-entry nudges before entry generated no interventions from P7 onward. Relational-meaning explanations (Figure~\ref{fig:mod-ba-b}) concentrated in P3's session, which motivated Modification 2, and recurred only once afterward (P7's question about edge thickness); P4's zero reflects his skipped network-graph tasks rather than resolved announcements. AT-intercepted shortcut failures (Figure~\ref{fig:mod-ba-c}) show the sharpest and most caveated pattern: they climb through P4 (6, all JAWS virtual cursor conflicts), and after Modification 3, no JAWS, NVDA, or magnification user recorded another one, but P5's seven failures on ZDSR, an assistive technology we had not previously tested, demonstrate that the fix was JAWS-specific rather than general. We read these panels as evidence that targeted modifications resolved the specific barriers they addressed for the AT configurations that surfaced them, while leaving open the configurations and barrier categories they did not target.

Next, we highlight three modifications that were particularly consequential for addressing the barriers above.

\paragraph{Modification 1: Explicit Entry Affordances for the Visualization Workspace} After P1's session, we identified that users could not reliably navigate into the main visualization area using the keyboard alone. P1 asked, "\textit{[How] do I go there?}" and required the facilitator to describe the visualization area's spatial location. We first trialed a single global shortcut (V) to jump focus into the workspace, but observed screen-reader conflicts with JAWS, NVDA, and ZDSR that made the shortcut inconsistently available; we therefore retired the shortcut and instead wrapped each visualization in a \texttt{FocusFeedbackWrapper} that exposes an explicit, labeled entry target for the first \texttt{Tab} stop, plays an entry earcon when focus enters the workspace, and plays an exit earcon when focus leaves it. In subsequent sessions, participants largely entered visualization areas without facilitator guidance (Figure~\ref{fig:mod-ba-a}); P6 required three re-entry nudges, and by P7, entering the visualization no longer generated an intervention. This modification illustrates that non-linear visualization content requires \textit{declared} entry affordances; Tab-order traversal alone is insufficient when the visualization is not a conventional document element, and "invisible" keyboard shortcuts can themselves become a barrier when they collide with AT input handling.

\paragraph{Modification 2: Directional, Weighted Connection Announcements} P3's session revealed that structural announcements in network graphs provided node names and connection counts but omitted the relational meaning of edges. P6 later echoed this: the system was "\textit{just telling [them] frustration connected to adaptation, resilience}" without conveying \textit{how} or \textit{how strongly} those codes were related. We modified the screen-reader output to split a node's connections into explicit \textit{outgoing} and \textit{incoming} lists (switchable with Up/Down arrows), and to attach a weight value and an importance qualifier (e.g., "strongest," "notable") to each edge, ordered by weight so the most structurally important connections are announced first. In later sessions, P6 and P7 both referenced edge direction and relative strength during network graph tasks (Figure~\ref{fig:mod-ba-b}), suggesting the enriched announcements made relational structure accessible without visual reference. This modification illustrates that non-hierarchical qualitative visualizations require representational strategies beyond node-and-edge enumeration: the screen-reader surface has to encode \textit{which way} an edge points and \textit{how much it matters}, not only \textit{that it exists}.

\paragraph{Modification 3: Replacing AT-Colliding Shortcuts with Persistent Labeled Buttons} P4's session exposed a recurring conflict between JAWS's virtual cursor and QUARTZ's global keyboard shortcuts: the virtual cursor intercepted keys such as \texttt{V}, \texttt{W}, and \texttt{S}, making them inconsistently responsive. P4 conditioned their adoption on resolution: "\textit{if the virtual cursor issue is resolved, meaning that with the virtual cursor on, I can use the keyboard shortcuts.}" Rather than fight the virtual cursor with a custom interception layer, an approach we found to be fragile across AT/browser combinations, we removed the colliding global shortcuts entirely and replaced them with a persistent, left-to-right toolbar of labeled buttons (\textit{Why This Type?}~$\rightarrow$~\textit{Evaluation}~$\rightarrow$~\textit{Next Visualization}~$\rightarrow$), keeping only in-visualization arrow-key navigation, which is activated once focus is inside the workspace and therefore does not collide with the virtual cursor. After this change, no JAWS, NVDA, or magnification user recorded another AT-intercepted shortcut failure (Figure~\ref{fig:mod-ba-c}), though P5's ZDSR conflicts show the fix did not generalize to untested screen readers. This modification illustrates that AT compatibility cannot always be achieved by \textit{negotiating with} assistive technologies. Sometimes, the more robust path is to \textit{design around} them, exposing the same affordances through interaction surfaces that the AT was built to handle well in the first place.
\section{Discussion}
\label{sec:discussion}

Our findings provide an empirical account of how BLV users select, create, navigate, and evaluate qualitative data visualizations. We synthesize across all four RQs and the cross-cutting barrier analysis to derive implications for the accessible visualization field.

\subsection{Qualitative Visualizations Demand Their Own Accessibility Infrastructure}
\label{subsec:disc_distinct_challenges}

A central finding is that the interaction paradigms, feedback mechanisms, and evaluation strategies that serve quantitative charts do not transfer to qualitative visualizations. This reflects a categorical difference, not degree.

RQ1 revealed that visualization selection scaffolding must contend with the interpretive subjectivity of qualitative representations. In quantitative visualization, the mapping from data to chart type is relatively constrained. In qualitative visualization, the same dataset can be rendered as a concept map, a network graph, or a Sankey diagram depending on the researcher's analytic goals. P5's response, that the system "\textit{made 99\% of the decision for [them]}", captures a tension that does not arise in quantitative chart recommendation: qualitative researchers expect agency over representational choices because those choices are themselves analytic acts. Static scaffolding that explains tradeoffs (as QUARTZ provides) is necessary but insufficient; participants like P5 wanted conversational, interactive explanation.

RQ3 demonstrated that the three-channel modality architecture (\textit{DG4}) functioned differently across visualization types in ways that have no parallel in quantitative chart accessibility. Sonification was most valued for Sankey diagrams, where P6 described it as enabling perception of the visualization's \textit{shape}, "\textit{you can easily go through the diagram and get a picture of what it might look like}", rather than simply encoding numerical values. Screen reader descriptions were essential for orientation in concept maps but insufficient for network graphs, where P6 found that listing connections without spatial context did not support interactive exploration. In quantitative chart accessibility, a single sonification-plus-text approach can enable access across bar charts, scatter plots, and line graphs~\cite{seoMAIDRMakingStatistical2024}. Our findings demonstrate that qualitative visualizations require modality-to-type tuning that the quantitative paradigm does not demand.

The cross-cutting barriers (Section~\ref{subsec:findings_barriers}) confirm what the related work anticipated: non-linear structures resist sequential traversal, semantic relationships require different representational strategies than numerical encodings, and AT diversity produces inconsistent experiences that heuristic evaluation cannot catch. P6's observation that concept maps, Sankey diagrams, and coding stripes allowed for more seamless, interactive exploration while network graphs did not highlights an additional challenge: even within the domain of qualitative visualization, each type's topology demands a distinct navigation model. A system that makes concept maps accessible does not, by extension, make network graphs accessible.

This finding aligns with \citet{zongRichScreenReader2022}'s observation that screen readers impose a linear interaction model on non-linear content, but extends it to a domain where the tension is amplified. A multi-series line chart at least has temporal or ordinal axes that provide sequential anchoring; a network graph has none.

\subsection{Design Implications for Accessible Qualitative Visualization}
\label{subsec:disc_design_guidelines}

We derive the following design implications (DIs) from our findings, grounded in our observations of our participants across four visualization types; therefore, we note that they are a starting point, \textit{not} a comprehensive standard.

\paragraph{DI1: Provide topology-aware navigation, not just keyboard access.} Keyboard accessibility is necessary but insufficient. Our structural navigation barriers (Section~\ref{subsec:findings_barriers}) showed that focus management failures and DOM-order traversal disoriented participants in graph-based visualizations. P6's contrast between concept maps (where "I'm able to press up and down and explore the diagram") and network graphs (where the system "just tells me frustration connected to adaptation") demonstrates that navigation must follow each visualization type's relational topology. This extends existing keyboard navigation guidelines~\cite{sharifVoxLensMakingOnline2022, elavskyDataNavigatorAccessibilityCentered2023}, which assume linear axis-based structures, to the non-linear topologies of qualitative data.

\paragraph{DI2: Map sonification to shape, not just magnitude.} P6's account of Sankey diagram sonification, hearing flow width through pitch rather than reading numbers, demonstrates that sonification in qualitative visualization should convey spatial and relational properties, not just numerical values. Where quantitative sonification maps pitch to data value~\cite{seoMAIDRMakingStatistical2024}, qualitative sonification should map sonic parameters to properties like flow width, connection density, and cluster membership. P6's report that sonification gave them "more confidence that it's going to look the way [they] wanted" suggests that this mapping supports not only comprehension but independent quality assessment (\textit{DG2}).

\paragraph{DI3: Scaffold the vocabulary, not just the interaction.} P7 and P8 identified unfamiliar terminology ("nodes," "edges") as their primary barrier, a conceptual challenge compared to quantitative, where "bar," "line," and "pie" are widely understood metaphors. Accessible qualitative visualization systems must scaffold the representational vocabulary of qualitative data alongside the interaction mechanics. \textit{DG1's} recommendation scaffolding partially addresses this, but our findings suggest that in-context definitions and examples during exploration, not only during selection, are needed.

\paragraph{DI4: Explain quality indicators, don't just display them.} P6's questioning of readiness percentages ("\textit{I wish they gave a more specific rubric}") and P8's request for a "quick guide" to understand detected limitations demonstrate that BLV users do not accept quality metrics at face value. Given the documented distrust of specialized tools~\cite{khan_i_2026}, quality indicators require contextual explanation, what was measured, why it matters, and what the user can do about it. This extends \textit{DG2} and \textit{DG3} beyond metric provision to metric interpretation.

\paragraph{DI5: Design for AT diversity from the outset.} The JAWS virtual cursor conflict that conditioned P4's adoption, the ZDSR rendering failures P5 experienced, and the narrator conflicts P2 reported were not edge cases, they affected participants using the three most common AT configurations and one less common one. Designing for one screen reader and assuming generalization is insufficient. This is consistent with prior work~\cite{seoMAIDRMakingStatistical2024, fanAccessibilityDataVisualizations2023} but particularly acute for qualitative visualizations, whose non-standard DOM structures stress-test ARIA implementations more aggressively than linear charts.

\paragraph{DI6: Balance scaffolding with agency.} P5's tension, valuing guidance but resisting automation, is a design challenge specific to qualitative visualization exploration, where representational choices are interpretive acts. Systems should scaffold decision-making through transparent explanation (as P2 valued) without foreclosing the researcher's agency over their own analytic representations. P5's suggestion of a conversational interface ("a small chat box where I can type questions") points toward interactive scaffolding as a potential resolution.

\subsection{Iterative Methods for Uncharted Design Spaces}
\label{subsec:disc_rite}

The RITE methodology was key in addressing participant feedback throughout the entire study. When designing accessible implementations of a novel visualization type, the space of possible barriers is too large and too dependent on AT diversity to be addressed through expert review. The trajectory we observed, from sessions requiring substantial facilitator intervention to sessions in which participants navigated with greater independence, demonstrates that early barriers were addressable through targeted modifications rather than indicative of fundamental design flaws.

RITE's value for accessibility research is how it produces design knowledge \textit{during} the evaluation rather than deferring it to a subsequent cycle. The guidelines above emerged from real interactions with real AT configurations, JAWS's virtual cursor behavior, ZDSR's rendering quirks, the specific ways P6 navigated Sankey diagrams versus concept maps. For researchers entering other uncharted accessible design spaces, RITE offers a principled way to converge on effective interaction patterns and reaffirms that upfront design cannot anticipate the full diversity of BLV users' needs.

\subsection{Limitations and Future Directions}
\label{subsec:limitations_and_future_directions}

This work provides valuable insights into accessible qualitative visualization design, though its limitations point directly to future directions. Our study included 8 BLV participants, consistent with prior evaluations of accessible visualization offerings~\cite{seoMAIDRMeetsAI2024, zongUmweltAccessibleStructured2024, blancoOlliExtensibleVisualization2022}. While appropriate for a formative RITE study, this sample does not support claims of generalizability; rather, its findings inform the design of accessible qualitative visualizations, whether those tested in this study or other types. Because the system changed between sessions, later performance reflects both system improvement and individual differences, and we analyzed these patterns as design trajectory rather than controlled experimental evidence. A subsequent summative evaluation, holding the system fixed across a larger, stratified sample (e.g., early- vs.\ late-onset vision loss, AT configuration, prior qualitative-research experience), would complement the formative insights reported here.
Our evaluation exercised both creation and exploration, but creation tasks used pre-structured data rather than participants' own research projects. Future work should evaluate the full authoring workflow, from raw qualitative data to finished visualization, in a more ecologically valid context. This includes longitudinal deployment studies with BLV researchers using QUARTZ and comparable accessible tools on their own data, which would provide stronger evidence of practical value and surface barriers that emerge only with sustained use in mixed-ability research teams~\cite{bennettInterdependenceFrameAssistive2018}. Such deployments are also the right setting in which to probe the \textit{epistemic verification barrier} we identified: P5's concern that non-visual output may not faithfully mirror what a sighted collaborator sees, and to test whether that barrier is resolved through collaborative verification, repeated exposure to a stable multimodal vocabulary, or both.
Our study also did not capture the full range of ATs that may be used with qualitative visualizations. The conflicts we observed with JAWS, NVDA, and ZDSR confirm that accessible qualitative visualization cannot be validated against a single AT configuration. Future work should evaluate a broader distribution of setups, including
refreshable Braille displays and tactile graphics, whose pairing with
conversational agents~\cite{reinders_when_2025} and whose tradeoffs against sonification~\cite{chundury_sound_2026} are now empirically characterized for quantitative data, as
well as mobile screen readers, and should do so in parallel with the
topology-specific navigation models our findings motivate: the tree-oriented paradigm that worked for concept maps did not transfer to network graphs, and we expect similar asymmetries to recur as new visualization types are added.
Finally, QUARTZ supports four qualitative visualization types; other qualitative displays, such as affinity diagrams, thematic maps, and journey maps, were not included and may present distinct challenges. Our findings around vocabulary barriers and the limited trust participants placed in unexplained readiness scores further point to two adjacent directions worth pursuing together: (i)~\textit{conceptual scaffolding} for participants new to qualitative-visualization literacy, including previewable exemplars and the conversational explanation P5 explicitly requested; and (ii)~\textit{explainable quality indicators} whose numerical outputs are backed by inspectable rubrics, so that BLV researchers can exercise informed agency over system recommendations rather than defer to the system when they cannot interrogate them.

\section{Conclusion}
\label{sec:conclusion}

We presented QUARTZ, a web-based system that provides multimodal, accessible representations of qualitative data visualizations and their evaluation through a RITE-driven study with 8 BLV participants across 12 tasks and four visualization types. Our findings expose accessibility challenges that are distinct from those in quantitative charts -- non-linear relational structures that resist sequential traversal, semantic relationships that demand different representational strategies than numerical encodings, and assistive technology diversity that no single-configuration test can capture. Through iterative co-design, we surfaced these barriers, resolved them, and documented the resulting design knowledge as guidelines for a visualization domain with limited empirical attention. We give a formative empirical answer to a foundational question: whether BLV users can independently navigate, comprehend, and derive meaning from qualitative data visualizations through accessible offerings. Our evidence indicates they can, when the tools are designed with them rather than retrofitted after the fact, and it maps the barriers that remain between this trajectory and a validated system. We seek to ensure that BLV researchers can engage with qualitative visualization as autonomous agents in their own research, not as recipients of access mediated by others.

\begin{acks}
  We thank all of our participants for their time and feedback on QUARTZ as we look to make qualitative visualizations more accessible for BLV researchers and beyond. We also thank all of our community partners for their support in distributing study materials. 
\end{acks}

\bibliographystyle{ACM-Reference-Format}
\bibliography{references/extra,references/a11y_framework, references/references}

\newpage
\appendix
\clearpage

\section{Implementation Details}
\label{appendix:implementation-details}

\subsection{Detection Layer}
\label{appendix:detection-layer-table}

Table~\ref{tab:detection-layer} documents the complete parameter set of the rule-based column-pattern analyzer: keyword weights, per-structure keyword sets, scoring contributions, structural bonuses, and the confidence gate. All values are fixed constants in \texttt{DataStructureAnalyzer}; no generative model is involved at any stage.

\begin{table}[tbhp]
\centering
  \caption{Rule-based column-pattern detection in \textsc{QUARTZ}
    (\texttt{DataStructureAnalyzer}): keyword weights, scoring
    contributions, and confidence gate. No generative model is used.}
  \label{tab:detection-layer}
  \Description{Reference table of the rule-based detection layer's fixed parameters, in five groups. Keyword weights: parent and child weigh 0.9; source and target 0.8; theme, category, from, and to 0.7; subtheme, subcategory, weight, and quantity 0.6; id 0.5; name and label 0.4; unlisted matcher tokens fall back to 0.5. Keyword sets: hierarchical (parent, child, theme, category, subtheme, subcategory), network (source, target, from, to, node1, node2), and flow (from, to, weight, quantity, amount, value). Keyword contributions: matched weights are summed and multiplied by 0.3 for hierarchical and flow patterns and 0.4 for network patterns. Structural bonuses: hierarchical adds (0.8 minus ratio) times 0.5 when the parent/child unique-ratio test passes; network adds ratio times 0.3 when source-target unique-count similarity exceeds 0.7; flow adds 0.2 when any numeric column is present. Aggregation: per-pattern scores are clamped to the 0-1 range, the highest-scoring structure is selected, and results below the 0.3 minimum confidence threshold are reported as unknown.}
  \centering
  \small
  \setlength{\tabcolsep}{4pt}
  \begin{tabular}{@{}llp{7cm}@{}}
    \toprule
    \textbf{Component} & \textbf{Parameter} & \textbf{Value / rule} \\
    \midrule
    Keyword weights
      & \texttt{parent}, \texttt{child} & $0.9$ \\
      & \texttt{source}, \texttt{target} & $0.8$ \\
      & \texttt{theme}, \texttt{category}, \texttt{from}, \texttt{to} & $0.7$ \\
      & \texttt{subtheme}, \texttt{subcategory}, \texttt{weight}, \texttt{quantity} & $0.6$ \\
      & \texttt{id} & $0.5$ \\
      & \texttt{name}, \texttt{label} & $0.4$ \\
      & Unlisted matcher tokens (\texttt{node1}, \texttt{node2}, \texttt{amount}, \texttt{value}) & fallback $0.5$ \\
    \midrule
    Keyword sets
      & Hierarchical & \texttt{parent}, \texttt{child}, \texttt{theme}, \texttt{category}, \texttt{subtheme}, \texttt{subcategory} \\
      & Network & \texttt{source}, \texttt{target}, \texttt{from}, \texttt{to}, \texttt{node1}, \texttt{node2} \\
      & Flow & \texttt{from}, \texttt{to}, \texttt{weight}, \texttt{quantity}, \texttt{amount}, \texttt{value} \\
    \midrule
    Keyword contribution
      & Hierarchical / Flow & $\sum$ matched weights $\times 0.3$ \\
      & Network & $\sum$ matched weights $\times 0.4$ \\
    \midrule
    Structural bonuses
      & Hierarchical & Parent/child unique-ratio test: if $\textit{ratio}<0.8$ and parent unique-ratio $<0.5$, add $(0.8-\textit{ratio})\times 0.5$ \\
      & Network & Source--target unique-count similarity: if $\textit{ratio}>0.7$, add $\textit{ratio}\times 0.3$ \\
      & Flow & Any numeric column present: $+0.2$ \\
    \midrule
    Aggregation
      & Per-pattern score & Clamped to $[0,1]$ \\
      & Selected structure & $\arg\max\{\text{hier.},\text{network},\text{flow}\}$ \\
      & Confidence gate & If best score $<0.3$ (\texttt{minConfidenceThreshold}), report \texttt{unknown} \\
    \bottomrule
  \end{tabular}
\end{table}

\clearpage

\subsection{Visualization Readiness Calculation}
\label{appendix:viz-readiness}

Table~\ref{tab:readiness-allocations} lists the additive point allocations the readiness calculator applies per visualization type. Each type is scored independently against its required and optional fields, capped at 100, and marked ready at or above 60.

\begin{table}[H]
  \centering
  \caption{Visualization readiness point allocations
    (\texttt{VisualizationReadinessCalculator}).
    Each visualization type is scored independently as an additive sum
    capped at $100$; a type is marked ready iff score $\geq 60$.}
  \label{tab:readiness-allocations}
  \Description{Point-allocation table for visualization readiness, one block per visualization type, each with a ready threshold of 60 or more points out of 100. Network graph: source-target mapping or parsed nodes and links (40), row count above zero (20), node labels (15), edge weights (15), relationship types (10). Concept map: parent-child mapping or hierarchical link types (40), content or descriptions (20), row count above zero (20), hierarchy level column (10), root nodes present (10). Sankey diagram: from-to mapping or flow links (40), flow values or weights (30), row count above zero (20), category column (10). Coding stripes: codes present (30), content or excerpts (30), source documents identified (20), character positions (20).}
  \centering
  \small
  \begin{tabular}{@{}llp{1cm}@{}} 
    \toprule
    \textbf{Visualization} & \textbf{Criterion} & \textbf{Pts} \\
    \midrule
    Network Graph
      & Source--target mapping or parsed nodes+links & $+40$ \\
      & Row count $>0$ & $+20$ \\
      & Node labels available & $+15$ \\
      & Edge weights available & $+15$ \\
      & Relationship types available & $+10$ \\
      & \textit{Ready threshold} & $\geq 60$ \\
    \midrule
    Concept Map
      & Parent--child mapping or hierarchical link types & $+40$ \\
      & Content/descriptions available & $+20$ \\
      & Row count $>0$ & $+20$ \\
      & Hierarchy level column & $+10$ \\
      & Root nodes present & $+10$ \\
      & \textit{Ready threshold} & $\geq 60$ \\
    \midrule
    Sankey Diagram
      & From--to mapping or flow links & $+40$ \\
      & Flow values/weights available & $+30$ \\
      & Row count $>0$ & $+20$ \\
      & Category column & $+10$ \\
      & \textit{Ready threshold} & $\geq 60$ \\
    \midrule
    Coding Stripes
      & Codes present & $+30$ \\
      & Content/excerpts available & $+30$ \\
      & Source documents identified & $+20$ \\
      & Character positions available & $+20$ \\
      & \textit{Ready threshold} & $\geq 60$ \\
    \bottomrule
  \end{tabular}
\end{table}


\section{User Study Materials}

\subsection{Task Questions}
\label{subsec:tasks-appendix}

\subsubsection{Concept Map}
\label{subsubsec:concept-map-qs}

\begin{enumerate}[label=\Roman*)]
    \item Import the provided CSV file into QUARTZ and talk me through your thoughts as you navigate it. Stop once you reach Step 6 of 7, or the "Visualization Readiness" page.
    \begin{enumerate}[label=\Alph*)]
        \item Explore the "Visualization Readiness" page. As you do, talk through your thoughts about its presentation. 
    \end{enumerate}

    \item Create two cross-links between two distinct pairs of  concepts of your choosing. Concepts should be unrelated, meaning they should not share a parent-child relationship.  

    \item Explore the "Evaluate" and "Why this type" panels to learn more about this visualization. As you familiarize yourself with each panel, talk through your thoughts about its content and layout. 
\end{enumerate}

\subsubsection{Network Graph}
\label{subsubsec:network-graph-qs}

\begin{enumerate}[label=\Roman*)]
    \item Explore the given network graph and describe its structure. 

    \item Update the network graph using the "Display" menu at the top of the visualization panel. Use any of the provided filters (e.g., node importance, edge weight, etc.), and let me know when you are satisfied with your changes.  

    \item Explore the "Evaluate" and "Why this type" panels to learn more about this visualization. As you familiarize yourself with each panel, talk through your thoughts about its content and layout. 
\end{enumerate}

\subsubsection{Sankey Diagram}
\label{subsubsec:sankey-qs}

\begin{enumerate}[label=\Roman*)]
    \item Explore the given Sankey diagram and describe its structure. 

    \item Create three annotations of your choosing using the Annotation Panel. Annotations can be about concepts or flows. Once complete, verify that you have created all three annotations. 

    \item Explore the "Evaluate" and "Why this type" panels to learn more about this visualization. As you familiarize yourself with each panel, talk through your thoughts about its content and layout. 
\end{enumerate}

\subsubsection{Coding Stripes}
\label{subsubsec:coding-qs}

\begin{enumerate}[label=\Roman*)]
    \item Explore the transcript and assess QUARTZ's recommendation for choosing coding stripes for this data. 

    \item Use the Code Bar to toggle on at least \textit{four} codes of your choosing. Afterwards, apply one or more codes of your choosing to any unlabeled text/data segment. Verify that you have applied the code(s) once complete.

    \item Find a segment of the data where multiple codes overlap; describe the codes present, their relative density, and their co-location. 
\end{enumerate}

\subsection{Post-Task Interview Protocol}
\label{subsec:interview-qs}

\begin{itemize}
    \item[$\star$] What was the most valuable aspect of using QUARTZ for you?
    
    \item[$\star$] What was the most challenging or frustrating part of the experience?
    
    \item[$\star$] How likely would you be to use QUARTZ in your actual research workflow? And why or why not?
    
    \item[$\star$] What features or improvements would make QUARTZ more useful for your needs?
    
    \item[$\star$] Is there anything else you would like to share about your experience?

\end{itemize}

\subsection{Accessibility Usability Scale (AUS)}
\label{subsec:aus-appendix}

\begin{enumerate}
    \item I think that I would like to use this system frequently for creating qualitative visualizations. [\textit{SD / D / N / A / SA}] 
    
    \item I found this system unnecessarily complex. [\textit{SD / D / N / A / SA}]
    
    \item I thought this system was easy to use. [\textit{SD / D / N / A / SA}]
    
    \item I think that I would need the support of a technical person to be able to use this system. [\textit{SD / D / N / A / SA}]
    
    \item I found the various functions in this system were well integrated. [\textit{SD / D / N / A / SA}]
    
    \item I thought there was too much inconsistency in this system. [\textit{SD / D / N / A / SA}]
    
    \item I would imagine that most BLV researchers would learn to use this system very quickly. [\textit{SD / D / N / A / SA}]
    
    \item I found this system very cumbersome to use. [\textit{SD / D / N / A / SA}]
    
    \item I felt very confident using this system. [\textit{SD / D / N / A / SA}]
    
    \item I needed to learn a lot of things before I could get going with this system. [\textit{SD / D / N / A / SA}]

\end{enumerate}

\subsection{RITE Iterations}
\label{subsec:rite-appendix}

\begin{description}
  \item[After P1] \textbf{Barrier:} Could not tell how QUARTZ interpreted the imported dataset or why a visualization type was recommended before committing, limiting informed selection.\\
  \textbf{Modification:} Added a persistent Data Interpretation Panel summarizing detected codes, documents, and participants with a plain-language rationale for each recommendation.\\
  \textbf{Rationale:} Give BLV users the evidence to match a visualization to their data non-visually instead of relying on sighted confirmation. (\textit{DG1, DG3})

  \item[After P2] \textbf{Barrier:} No keyboard-only path to select text for coding; the \texttt{D}-key codebook flickered open/closed; auxiliary buttons created extraneous focus stops.\\
  \textbf{Modification:} Added a Shift{+}Arrow text-selection hook with screen-reader announcement of the current selection, stabilized the codebook toggle, and removed non-essential buttons.\\
  \textbf{Rationale:} Make the core coding workflow fully operable without a mouse while reducing cognitive load. (\textit{DG2, DG4})

  \item[After P3] \textbf{Barrier:} Built-in narration in the network graph conflicted with the participant's own screen reader, masking sonification; Sankey contrast and \texttt{V}-key focus behavior were inconsistent.\\
  \textbf{Modification:} Removed the built-in narration layer, unified \texttt{V}-key focus isolation across visualizations, and raised Sankey contrast to meet WCAG~AA.\\
  \textbf{Rationale:} Let participants use their preferred AT without dueling speech and keep multi-modal feedback consistent. (\textit{DG4})

  \item[After P4] \textbf{Barrier:} Sonification pitch range and dead-end cues were hard to distinguish; screen-reader focus leaked out of the evaluation modal into the page behind it.\\
  \textbf{Modification:} Re-tuned the network sonification (pitch, dead-end earcon, glissando), added a Sonification Guide, and fixed tab order and dialog focus management for the evaluation modal.\\
  \textbf{Rationale:} Keep audio feedback legible and scope modal interaction to the active dialog. (\textit{DG4})

  \item[After P5] \textbf{Barrier:} Did not discover keyboard shortcuts for switching visualizations or triggering \textit{Evaluate}/\textit{Why This Type?}; hidden affordances slowed task progress.\\
  \textbf{Modification:} Replaced those shortcuts with always-visible buttons in a fixed bottom toolbar ordered \textit{Why This Type?}~$\rightarrow$~\textit{Evaluation}~$\rightarrow$~\textit{Next Visualization}~$\rightarrow$.\\
  \textbf{Rationale:} Make the task flow explicit rather than relying on memorized shortcut discovery. (\textit{DG2})

  \item[After P6] \textbf{Barrier:} Sidebar added extraneous focus stops; first Tab onto the network canvas did not activate arrow-key navigation; number-key labels in the codebook implied shortcuts that did not actually assign codes.\\
  \textbf{Modification:} Hid the sidebar behind \textit{Next Visualization}, wired the first Tab to enter arrow-key navigation with a spoken summary, and removed the misleading number-key labels.\\
  \textbf{Rationale:} Minimize focus noise and align keyboard affordances with actual behavior. (\textit{DG2, DG4})

  \item[After P7] \textbf{Barrier:} Screen reader listed a node's connections as a flat list, hiding structural importance; modal focus occasionally restarted at the page top on some AT/browser pairs.\\
  \textbf{Modification:} Ordered connection announcements by degree/centrality with explicit importance phrasing, and hardened the focus trap across all modals.\\
  \textbf{Rationale:} Convey graph structure non-visually and keep interaction bounded to the active context. (\textit{DG3, DG4})

  \item[After P8] \textbf{Barrier:} Residual cross-AT inconsistencies: eval-modal tabbing still leaked on some JAWS/NVDA+browser pairs, selection announcements differed across AT, and the keyboard contract was implicit.\\
  \textbf{Modification:} Codified a \texttt{networkGraph\-Keyboard\-Contract} with contract tests, added unit tests for selection announcements, an end-to-end modal-focus-trap suite, and patched remaining eval-modal tabbing paths.\\
  \textbf{Rationale:} Lock accessibility behavior into executable specifications so prior RITE fixes remain stable across AT. (\textit{DG2, DG4})
\end{description}

\end{document}